\documentclass[prd,aps,twocolumn,nofootinbib,superscriptaddress,eqsecnum,floatfix,preprintnumbers,amsmath,amssymb,
nofootinbib,longbibliography,]{revtex4-1}

\usepackage{amssymb,amsmath}
\usepackage{epsfig}
\usepackage[dvipsnames]{xcolor}
\usepackage[utf8]{inputenc}
\usepackage{stmaryrd}
\usepackage{mathrsfs}
\usepackage{mathalfa}
\usepackage{accents}
\usepackage[normalem]{ulem}

\newcommand{\dd}{\mathrm{d}}
\newcommand{\ii}{i}

\newcommand{\be}{\begin{equation}}
\newcommand{\ee}{\end{equation}}
\newcommand{\bea}{\begin{eqnarray}}
\newcommand{\eea}{\end{eqnarray}}
\newcommand{\ba}{\begin{equation}\begin{aligned}}
\newcommand{\ea}{\end{aligned}\end{equation}}

\newcommand{\beg}{\begin{gather*}}
\newcommand{\eng}{\end{gather*}}
\newcommand{\hh}{,\hspace{0.5cm}}
\newcommand{\hhh}{,\hspace{0.2cm}}

\newcommand{\eq}[1]{(\ref{#1})}

\newcommand{\n}[1]{\label{#1}}

\newcommand{\CAL}{\mathcal}

\newcommand{\ts}[1]{{\boldsymbol{#1}}}

\def\XXint#1#2#3{{\setbox0=\hbox{$#1{#2#3}{\int}$ }
\vcenter{\hbox{$#2#3$ }}\kern-.6\wd0}}

\usepackage[makeroom]{cancel}
\usepackage[caption=false]{subfig}
\usepackage[colorlinks=true,
            citecolor=green,
            linkcolor=red,
            filecolor=cyan,
            urlcolor=magenta,
            backref=false]{hyperref}

\begin{document}

\title{Spherically symmetric black holes in the limiting curvature theory of gravity}

\author{Valeri P. Frolov}
\email{vfrolov@ualberta.ca}
\affiliation{Theoretical Physics Institute, University of Alberta, Edmonton, Alberta, Canada T6G 2E1}
\author{Andrei Zelnikov}
\email{zelnikov@ualberta.ca}
\affiliation{Theoretical Physics Institute, University of Alberta, Edmonton, Alberta, Canada T6G 2E1}


\begin{abstract}
In this paper we describe a model of  a four-dimensional spherically symmetric black hole in a limiting curvature theory of gravity. In this theory the Einstein-Hilbert action is modified by adding to the action terms providing inequality constraints on chosen curvature invariants. We demonstrated that in such a model, instead of formation of the curvature singularity, the spacetime remains regular in the black hole interior. For black holes with gravitational radius much larger than the radius $\ell$ of the critical curvature the obtained solutions describe a space exponentially expanding in one direction and periodically oscillating in the other two (spherical) directions.
\medskip

\hfill {\scriptsize Alberta Thy 36-21}
\end{abstract}

\maketitle

\section{Introduction}

Existence of singularities in General Relativity (GR) is a well known ``headache" of this theory. According to GR they exist both in cosmology and inside black holes. Famous theorems proved by Penrose and Hawking imply that such singularities are inevitable provided physically reasonable energy conditions are satisfied. Exact solutions of Einstein equations show that both cosmological and black hole singularities are related to an infinite growth of spacetime curvature. In spite of this common property there exists an important difference between these two types of singularities. Homogeneous isotropic universes are conformally flat, so that their Weyl tensor vanishes. Their singularity is related to an infinite growth of the Ricci tensor. In order to prevent its formation it is sufficient to modify the matter stress-energy tensor. Certainly, such a modification should violate some energy conditions. Inside a static or stationary black hole the singularity is related to infinite growth of the Weyl tensor and it cannot be prevented by simple modification of the matter stress-energy tensor. This property makes the problem of black hole singularities more complicated.

There is a general belief that the ultraviolet (UV) incompleteness of GR can be cured by a properly chosen modification of this theory. Namely, one can expect that in such a modification the curvature invariants cannot grow infinitely and are always less than some limiting value. This condition was formulated by Markov \cite{Markov:1982,Markov:1984ii} as a new fundamental principle of theoretical physics.
A natural question is: If curvature singularities are eliminated in a theory satisfying the limiting curvature condition, how the spacetime properties would be modified in the domains with high curvature close to the critical one. There are several publications where this question was addressed. It was demonstrated that in cosmology the existence of the limiting curvature opens a possibility of bouncing cosmological models \cite{Markov:1982,Markov:1984ii,Gasperini:1992em,Gasperini:2002bn,Mukhanov:1991zn,Brandenberger:1993ef,Turok_2005,Biswas_2006,Barvinsky:2008ia,Novello:2008ra,Lehners:2008vx,Ashtekar_2009,CesareSilva:2020ihf,Biswas:2012bp,Battefeld:2014uga,Brandenberger_2017,Yoshida:2017swb,Ijjas_2018}. For black holes the validity of this principle makes it possible to form new universes in their interior \cite{Frolov:1988vj,Frolov:1989pf,Morgan:1990yy,Barrabes:1995nk}.

There exist several models which obey the limiting curvature condition. For example, one can include in the theory of gravity dilaton scalar fields and find their potential which allows one to restrict curvature (see e.g. \cite{Brandenberger_2017,Yoshida:2017swb,Muknanov:2017}).

Recently, the authors of the present paper proposed a new approach in which the limiting curvature condition is satisfied. This is achieved by adding to the Einstein-Hilbert action of GR extra terms which provide fulfilment of inequality constraint(s) restricting the value of chosen curvature invariant(s). Basic mechanism of application of such inequality constraints to classical mechanics is discussed in \cite{AVF}.  Limiting curvature models for 2D black holes and for cosmology were studied in \cite{Frolov:2021kcv} and \cite{frolov2021bouncing}, respectively.

In the present paper we continue to study the limiting curvature theory of gravity. Now we focus on the problem of the interior of a four-dimensional spherically symmetric black hole. In section \ref{SecII} we discuss spherically symmetric metrics and their curvature invariants. In section \ref{SecIII} we present a reduced action approach. Modification of the gravitational theory by including inequality constraint restricting curvature is described in section \ref{SecIV}. Linear in curvature invariants inequality constraints are discussed in section \ref{SecV}. Next four sections contain derivation and analysis of solutions describing the interior of the black hole in the constructed limiting curvature models. Summary of results and their discussion is the subject of section \ref{SecX}. An appendix collects useful formulas, which are used in the ``main body" of the paper.

\section{Spherically symmetric metric and its curvature invariants}\label{SecII}

\subsection{Metric}

Schwarzschild metric is a spherically symmetric solution of vacuum Einstein equations. It has a well known form
\be \n{a1}
ds^2=-(1-\frac{2M}{r} )dt^2+\frac{dr^2}{1-\frac{2M}{r}}+r^2 d\omega^2\, ,
\ee
where $d\omega^2=d\theta^2+\sin^2\!\theta\, d\phi^2$ is a metric on 2D round sphere. The existence of the Killing vector $\ts{\xi}=\partial_{t}$ for this metric is a consequence of Birkhoff's  theorem. The parameter $M$ is the mass of a static black hole described by  metric (\ref{a1}). In the black-hole's interior, that is for $r<2M$, we write this metric in the form
\be\n{a2}
ds^2=-\frac{dr^2}{f}+f dt^2+r^2 d\omega^2\hh f=\frac{2M}{r}-1 .
\ee
The coordinate $t$ is still the Killing parameter, but now $(\nabla t)^2>0$, so that $t$ is a spatial coordinate.
The metric function $f=-(\nabla r)^2$ is positive in the black hole interior, so that the coordinate $r$ is a timelike coordinate.
We define a new time coordinate $\tau$ as follows
\be
\tau =-\int_{2M}^r \frac{dr}{\sqrt{f}} .
\ee
Calculating this integral one finds the following relation between $\tau$ and $r$
\be
\tau=\sqrt{r(2M-r)}+M \arcsin\Big(\frac{M-r}{r}\Big)+\frac{1}{2}\pi M .
\ee
Parameter $\tau$  has a simple meaning. Namely, it is a proper time along a world line with fixed values of $(t,\theta,\phi)$. The integration constant is chosen so that $\tau=0$ at $r=2M$. At $r=0$ one has $\tau=\pi M$.
Kretschmann curvature invariant ${\CAL K}=R_{\alpha\beta\gamma\delta}R^{\alpha\beta\gamma\delta}$ calculated for this metric is
\be
{\CAL K}=\frac{48 M^2}{r^6} .
\ee
It infinitely grows in the black hole interior in the vicinity of a singularity at $r= 0$.

The existence of singularities is a well known "illness" of the General Relativity.
Our goal is to study how this prediction is changed if one modifies the theory by imposing the limiting curvature restriction. In such a model the metric (\ref{a2}) remains valid until the spacetime curvature reaches some limiting value. Later we specify its concrete choice but at the moment we assume that this happens at some value of the radial coordinate $r=r_0<2M$.

For the later time the form of the metric would be different from (\ref{a2}). We write this metric in the following general form
\ba\n{metric}
ds^2=-b^2 d\tau^2+B^2 dt^2+a^2 d\omega^2 \, ,
\ea
where $a$, $b$ and $B$ are functions of  time $\tau$. Let us make some explanation concerning this choice.
Metric (\ref{metric}) has the Killing vector $\ts{\xi}=\partial_t$ which it inherited from (\ref{a2}) and $t$ is a spacelike coordinate. There is an ambiguity in the choice of the time coordinate $\tau$ and the presence of the metric coefficient $b(\tau)$ reflects this. In a special gauge $b(\tau)=1$ the coordinate $\tau$ coincides with the proper time parameter. At the moment we keep $b(\tau)$ arbitrary. This allows one to reduce the action of the theory to a functional depending on three functions, $b(t)$, $a(\tau)$ and $B(\tau)$, such that its variations give a complete set of equations. After the variation one can fix the choice of gauge.

We shall use two options:
\begin{itemize}
\item Synchronous gauge: $b(\tau)=1$
\be \n{sm}
ds^2=-d\tau^2+B(\tau)^2 dt^2+a(\tau)^2 d\omega^2 .
\ee
\item Radial gauge in which
\be \n{rm}
ds^2=-\frac{1}{f(r)}\, dr^2+e^{2\gamma(r)}f(r) \,dt^2+r^2 d\omega^2 .
\ee
\end{itemize}
It is easy to see that
\be
\ts{\xi}^2=B^2=e^{2\gamma}f
\hh f=-(\nabla r)^2  .
\ee

Any symmetric tensor $A_{\mu\nu}$ which respects the symmetries of the metric (\ref{metric}), that is which is invariant under rotations,  translation $t\to t+a$ and reflection $t\to -t$, has the following form
\be \n{AA}
A_{\mu}^{\nu}=\mbox{diag}(\accentset{0}{\CAL A}(\tau),\accentset{1}{\CAL A}(\tau),\hat{\CAL A}(\tau),\hat{\CAL A}(\tau)) .
\ee

\subsection{Curvature invariants}

The Riemann curvature tensor for the metric (\ref{sm}) has four non-vanishing components\footnote{
This property is also valid for a spherically symmetric metrics which depends on both space and time coordinates.
Narlikar and Karmarkar \cite{NARLIKAR} proposed that this set of four invariants is sufficient for the construction of all other algebraic invariants of the curvature tensor. The metric (\ref{metric}) belongs to a wider class of so called warped product metrics. A discussion of the complete set of curvature invariants which is sufficient for construction of algebraic curvature invariants and further references can be found in \cite{CURV_INV_1998}.
}
\bea\n{RIN1}
&& R_{\hat{\tau}\hat{t}\hat{\tau}\hat{t}}=-v\hh R_{\hat{\tau}\hat{\theta}\hat{\tau}\hat{\theta}}=-q\, ,\\
&&R_{\hat{t}\hat{\theta}\hat{t}\hat{\theta}}=u, \hskip 0.77cm   R_{\hat{\theta}\hat{\phi}\hat{\theta}\hat{\phi}}=p .
\n{RIN2}
\eea
The hat over indices of the curvature means that the  components of this tensor are calculated in the tetrad formed by unit vectors along the corresponding coordinate lines.
The curvature invariants $(p,q,u,v)$ in synchronous coordinates are
\bea\n{pquv}
&&p=\frac{\dot{a}^2+b^2}{a^2b^2}, \hskip 0.7cm
q=\frac{\ddot{a}}{ab^2}-\frac{\dot{a}}{a}\frac{\dot{b}}{b^3},\\
&&u=\frac{\dot{a}\dot{B}}{a B b^2} , \hskip 1cm
v=\frac{\ddot{B}}{Bb^2}-\frac{\dot{B}}{B}\frac{\dot{b}}{b^3}  \n{uv}.
\eea
A dot in these expressions means the derivative with respect to $\tau$.
The curvature invariants in the radial coordinates are
\bea\label{f}
&p=\frac{f+1}{r^2}\hh q=\frac{{f'}}{2r}\hh u=\frac{f {\gamma'}}{r}+\frac{f'}{2r}\, ,\\ &v=f{\gamma'}+f{\gamma'}^2+\frac{1}{2}{f''}+\frac{3}{2}{f'}{\gamma'} .
\eea
Here the prime means a derivative with respect to $r$.

In what follows the following simple remarks will be useful.
\begin{itemize}
\item Let us consider a two dimensional spacetime with metric
\be \n{gg}
d\gamma^2=-b^2(\tau)+B^2(\tau) dt^2 .
\ee
Its 2D curvature  is ${}^{(2)}R=2 v$ , where $v$ is given in (\ref{uv}).
\item Consider three dimensional spacetime with metric
\be \n{GG}
d\Gamma^2=-b^2(\tau)d\tau^2+a(\tau)^2 d\omega^2 .
\ee
The Einstein tensor for it is ${}^{(3)}G^{\mu}_{\nu}=-\mbox{diag}(p,q,q)$,
where $p$ and $q$ are given in (\ref{pquv}).
\end{itemize}
Hence, if one restricts 2D curvature for metric $d\gamma^2$ and 3D curvature invariants  for metric $d\Gamma^2$ then the 4D curvature invariants (\ref{RIN1})- (\ref{RIN2}) will be also restricted, provided an additional invariant $u$ is bounded.

For the Schwarzschild metric these curvature invariants take the form
\be\n{TRAN}
p=v=-2q=-2u=\frac{2M}{ r^3} .
\ee
The Ricci tensor, the traceless Ricci tensor, and the Einstein tensor for the metric \eq{metric} are linear combinations of the basic invariants $(p,q,u,v)$. These expressions as well as expressions for quadratic in curvature invariants are collected in the appendix.

The curvature invariants $q$ and $v$ which enter the expression (\ref{RIN1}) for the Riemann tensor contain second derivatives of the metric functions $a(\tau)$ and $B(\tau)$.
Using covariant derivatives of the Riemann curvature one can construct additional set of scalar invariants, which contain higher derivatives of these functions. For example, one has
\bea\n{DRIN1}
&& R_{\hat{\tau}\hat{t}\hat{\tau}\hat{t} ;\hat{\tau}}=-\frac{a}{B}\dot{v}+\ldots \, ,\\
&&R_{\hat{\tau}\hat{\theta}\hat{\tau}\hat{\theta} ;\hat{\tau}}= R_{\hat{\tau}\hat{\phi}\hat{\tau}\hat{\phi} ;\hat{\tau}}=-\dot{q}+\ldots  .
\n{DRIN2}
\eea
As earlier we use the hat over indices to indicate that the  components of this tensor are calculated in the orthonormal tetrad. The dots denote terms which contain less than three time derivatives of the metric functions. All other components of the covariant derivatives of the Riemann tensor which contain $\dot{q}$ and $\dot{v}$ can be obtained by permutations of the indices in (\ref{DRIN1}) and (\ref{DRIN2}).

\section{Reduced action}\label{SecIII}

Let us consider an action
\ba\n{SSS}
S=\frac{1}{\kappa}\int \dd^4 x \,\sqrt{-g} L(g)\hh
\kappa=8\pi G .
\ea
Here the Lagrangian $L(g)$ depends on the metric $\ts{g}$ and its derivatives. Besides this it may depend also on some other variables which will be specified later.
We use the following notation for  the variation of this action over the metric
\ba\n{TT}
{T}^{\alpha\gamma}=\frac{2}{\sqrt{-g}}\frac{\delta S}{\delta g_{\alpha\gamma}} .
\ea
We assume that the Lagrangian $L$ is a scalar and  equations of motion for other fields which enter $L$ are satisfied then the following conservation law is valid
\ba
{\cal T}^{\alpha\beta}{}_{;\beta}=0 .
\ea

Tensor $\ts{T}$  being calculated for the metric (\ref{metric}) has the form (\ref{AA}).
We call such a tensor a reduced one.
This reduced  tensor can be obtained in a different way. Let us substitute first the ansatz (\ref{metric}) into the Lagrangian $L(g)$. As the result one gets the Lagrangian as a function of $b(\tau)$, $a(\tau)$, $B(\tau)$ and their derivatives.  Since it does not depend on $t$, $\theta$ and $\phi$, one can integrate the action $S$ over these coordinates. As a result one has
\be
S=\frac{1}{\kappa}V {\CAL S} ,
\ee
where we defined the dimensionally reduced action
\be
{\CAL S}= \int \dd\tau\, a^2 b B \, L(a,b,B)\, ,
\ee
and the volume
\ba
V=\int \sin\theta\, \dd\theta\dd\phi\dd x=4\pi\int \dd t .
\ea
This (formally infinite) factor does not affect the equations of motion derived from the action and can be omitted.

By varying the reduced action with respect to its arguments $b$, $a$ and $B$ one obtains three quantities, which coincide with the components of the reduced tensor $\ts{\CAL T}$
\bea
&&{\cal T}_{\tau}^{\tau}=\frac{1}{a^2B}\frac{\delta{\CAL S}}{\delta b},
\\
&&{\cal T}_{t}^{t}=\frac{1}{a^2 b}\frac{\delta {\CAL S}}{\delta B},
\\
&&{\cal T}_{\theta}^{\theta}={\cal T}_{\phi}^{\phi}=\frac{1}{2a b B}\frac{\delta{\CAL S}}{\delta a} .
\eea
In what follows we shall use the approach based on the reduced action for obtaining the gravitational equations in our model.

For the metric (\ref{metric}) the conservation law (\ref{TT}) gives the following relation
\ba\n{CONSERV}
&\partial_{\tau}{\cal T}_{\tau}^{\tau}=\frac{\dot{B}}{B}\big[{\cal T}_{t}^{t}-{\cal T}_{\tau}^{\tau}\big]
+2\frac{\dot{a}}{a}\big[{\cal T}_{\theta}^{\theta}-{\cal T}_{\tau}^{\tau}\big] .
\ea
Thus if we know ${\cal T}_{\tau}^{\tau}$ and ${\cal T}_{\theta}^{\theta}$ components, then ${\cal T}_{t}^{t}$  can be expressed in terms of them.

\section{Inequality constraints}\label{SecIV}

In a model with the limiting curvature we consider the action (\ref{SSS}) with the Lagrangian $L(g)$ of the form
\be
L(g)=L_g +L_{c} .
\ee
Here $L_g=\frac{1}{2}R$ is the Lagrangian for Einstein-Hilbert action and $L_{c}$ is a part of the Lagrangian responsible for inequality curvature constraints. In what follows we use ansatz (\ref{metric}) for the metric, and  our starting point is the following reduced action
\bea \n{action}
& {\CAL S}=\int d\tau a^2 b B \,{\CAL L}\, ,\\
&{\CAL L}={\CAL L}_g+\sum_j \chi_j (\Phi_j+\zeta_j^2) .\n{RedL}
\eea
Here ${\CAL L}_g$ is the reduced form of the Einstein-Hilbert action
\be
{\CAL L}_g=\frac{1}{2} R=p+2q+2u+v .
\ee
After integration by parts one can present this part of the action in the form
\be
{\CAL S}_g=\int d\tau B \left[b-\frac{\dot{a}^2}{ b}-2\frac{a\dot{a} \dot{B}}{Bb}\right] .
\ee
Its variations give
\bea
&&{\cal G}_{\tau}^{\tau}=\frac{1}{a^2B}\frac{\delta{\CAL S}_g}{\delta b}=-(p+2u),\\
&&{\cal G}_{t}^{t}=\frac{1}{a^2 b}\frac{\delta {\CAL S}_g}{\delta B}=-(p+2q),\\
&&{\cal G}_{\theta}^{\theta}=\frac{1}{2a bB}\frac{\delta{\CAL S}}{\delta a}
=-(q+u+v) .
\eea
It is easy to see that as expected the obtained expressions coincide with the components of the reduced Einstein tensor (see Appendix).

The action (\ref{action}) contains also a part which is responsible for the inequality constraints. We choose constraint functions in the form
\be
\Phi_j=\Phi_j(p,q,u,v;\Lambda_j) .
\ee
Basically, $\Phi_j$ is a scalar invariant constructed from the curvature tensor and $\Lambda_j$ is a limiting value parameter for this invariant. Each of such constraints is accompanied by a pair of Lagrange multipliers $\chi_j$ and $\zeta_j$. By varying the action ${\CAL S}$ over these multipliers one gets relations
\be
\Phi_j(p,q,u,v;\Lambda_j)+\zeta^2_j=0\hh \chi_j\zeta_j=0 .
\ee
The first of these relations shows that $\Phi_j\le 0$ so that $\zeta^2_j=-\Phi_j$. If $\Phi_j< 0$ the second relation implies that $\chi=0$. This function becomes nonzero only when the first equality is saturated and the curvature invariant $\Phi_j$ reaches its limiting value
\be
\Phi_j=0 .
\ee
We call $\chi_j(\tau)$ a control function.

When all control functions vanish, the constrains do not contribute into the gravity equations, so that they are identical with the equations obtained from the reduced Einstein-Hilbert action. We call this regime subcritical. After one of the constraints is saturated and the corresponding curvature invariant reaches its critical value a solution becomes supercritical. It follows along this constraint, while the gravitational equations will be modified by adding terms dependent on $\chi$ and its derivatives which can be obtained by variation  over the metric of the corresponding constraint term.

After a subcritical solution enters the supercritical regime it behavior may be different. If at some moment of time $\tau_1$ the control function $\chi_j(\tau_1)$ (with some of its derivatives) vanishes and the other constraint functions are still non-saturated, the solution can return to its subcritical regime. If during the supercritical evolution along the constraint number $j$ the other constraint function (say with number $i\ne j$) becomes saturated, the system can slip to it or its motions will be restricted by both constraints simultaneously. These options and other properties of solutions of the considered model with limiting curvature depend on the number of the inequality constraints and their structure.

\section{Linear constraints}\label{SecV}

In this paper we consider linear in curvature constraints. The most general form of such a constraint is
\be \n{LIN_C}
\Phi\equiv \rho p-\mu q+\nu u + \sigma v -\Lambda=0
\hh
\Lambda=\frac{1}{\ell^2} .
\ee
Here $p$, $q$, $u$ and $v$ are basic curvature invariants defined in (\ref{pquv}), and $\rho$, $\mu$, $\nu$, and $\sigma$ are dimensionless constants. Our choice of sign for the term with the coefficient $\mu$ will be convenient in further discussions. Parameter $\Lambda$ plays the role of the limiting curvature value. It has dimension [length]$^{-2}$. We denote by $\ell$ the corresponding critical length scale.

As we already mentioned the control function $\chi$ in the subcritical regime vanishes and the gravitational equations are identical with the standard Einstein equations. For our problem this means that the subcritical solution coincides with the Schwarzschild metric \eq{a1}.
Let us discuss the form of these equations in the supercritical regime.  The reduced Lagrangian (\ref{RedL}) at this phase is of the form
\be
{\CAL L}=(p+2q+2u+v)+(\rho p-\mu q+\nu u +\sigma v-\Lambda)\chi .
\ee
Variation with respect to control function $\chi$ give the constraint (\ref{LIN_C}), while the gravitational equations gives the following set of equations.
\begin{itemize}
\item The $(\tau,\tau)$ equation
\ba\label{E00}
\frac{\dot{a}^2}{a^2}+&\frac{1}{a^2}+2\frac{\dot{a}\dot{B}}{aB}\, \\
=&\rho \Big(
-\frac{\dot{a}^2}{a^2}+\frac{1}{a^2}
\Big)\chi -\mu\Big[
\frac{\dot{a}}{a}\dot{\chi}+\Big(\frac{\dot{a}^2}{a^2}+\frac{\dot{a}\dot{B}}{aB}
\Big)\chi
\Big]\, \\
&-\nu\frac{\dot{a}\dot{B}}{aB}\chi+\sigma \frac{\dot{B}}{B}\Big[\dot{\chi}+
2\frac{\dot{a}}{a}\chi \Big]-\Lambda\chi .
\ea
\item The $(t,t)$ equation
\ba\label{E11}
\frac{\dot{a}^2}{a^2}&+\frac{1}{a^2}+2\frac{\ddot{a}}{a}
=\rho \Big(\frac{\dot{a}^2}{a^2}+\frac{1}{a^2}
\Big)\chi -\mu\frac{\ddot{a}}{a} \chi\\
&-\nu\Big[\frac{\dot{a}}{a}\dot{\chi}+\Big(\frac{\ddot{a}}{a}+\frac{\dot{a}^2}{a^2}\Big)\chi\Big]\\
&+\sigma\Big[\ddot{\chi}+4\frac{\dot{a}}{a}\dot{\chi}+2\Big(\frac{\ddot{a}}{a}+\frac{\dot{a}^2}{a^2}\Big)\chi\Big] -\Lambda\chi .
\ea
\item The $(\theta,\theta)$ equation
\ba\label{E33}
\frac{\ddot{a}}{a}&+\frac{\dot{a}\dot{B}}{aB}+\frac{\ddot{B}}{B}
=-\rho \Big[\frac{\dot{a}}{a}\dot{\chi}+\Big(
\frac{\ddot{a}}{a}+\frac{\dot{a}\dot{B}}{aB}\Big)\chi\Big]\\
&- \frac{\mu}{2}\Big[ \ddot{\chi}
+\Big(2\frac{\dot{a}}{a}+2\frac{\dot{B}}{B}\Big)\dot{\chi}+\Big(
2\frac{\ddot{a}}{a}+2\frac{\dot{a}\dot{B}}{aB}+\frac{\ddot{B}}{B}\Big)\chi
\Big]\\
&-\frac{\nu}{2}\Big[\frac{\dot{B}}{B} \dot{\chi}+\frac{\ddot{B}}{B}\chi
\Big]+\sigma
\frac{\ddot{B}}{B}\chi-\Lambda \chi .
\ea
\end{itemize}

Thus we have a set of four equations (\ref{LIN_C}), (\ref{E00}), (\ref{E11}) and (\ref{E33}) for three functions $a(\tau)$, $b(\tau)$ and $\chi(\tau)$. However, they are not independent. The conservation law (\ref{CONSERV})
establishes relation between equations (\ref{E00})-- (\ref{E33}). One can  reduce  the total number of independent equations for three functions $a(\tau)$, $b(\tau)$,  $\chi(\tau)$ to two in two different ways:
\begin{itemize}
\item Use besides the constraint (\ref{LIN_C}) the first order equation (\ref{E00}) together with one of the other two equations,  (\ref{E11}) and (\ref{E33});
\item Use equations  (\ref{LIN_C}), (\ref{E11}) and (\ref{E33}).
\end{itemize}
In the latter case the equation  (\ref{E00}) should be imposed on the initial conditions at some time $\tau_0$. The conservation law  (\ref{CONSERV}) guarantees that the system is consistent and condition  (\ref{E00}) is valid for any $\tau$.

Let us consider the first option.
Using the constraint equations we write the corresponding  three equations containing second derivatives
in the following form
 \ba
-\mu q+\sigma v=\Lambda-\rho p-\nu u ,
\ea
\ba\label{E11a}
[2+&(\nu-2\sigma)\chi]q+\sigma v-\sigma \ddot{\chi}\\
=&-p
-\nu\Big[\frac{\dot{a}}{a}\dot{\chi}+\Big(p+u-\frac{1}{a^2}\Big)\chi\Big]\\
&+\sigma\Big[4\frac{\dot{a}}{a}\dot{\chi}+2\Big(p-\frac{1}{a^2}\Big)\chi\Big] ,
\ea
\ba\label{E33b}
(1+\rho \chi)q&+[1+\frac{1}{2}(\mu+\nu)]\chi v+\frac{1}{2}\mu \ddot{\chi}\\
=&-u-\Big[(\rho+\mu)\frac{\dot{a}}{a}+\Big(\mu+\frac{\nu}{2}\Big)\frac{\dot{B}}{B}\Big]\dot{\chi}\\
&-\rho(p+u)\chi-(\mu+\nu)u \chi .
\ea
This system of equations is written in such a form where the terms with second derivatives stand in their left-hand side, while the right-hand side does not contain these second order derivatives.
This set of equations  can be written in a matrix form
\begin{equation*}
U
\begin{pmatrix}
q \\
v  \\
\ddot{\chi}
\end{pmatrix}=\dots ,
\end{equation*}
where $U$ is the following $3\times 3$ matrix
\begin{equation*}
U=
\begin{pmatrix}
-\mu & \sigma & 0  \\
2+(\nu-2\sigma)\chi & \sigma\chi &-\sigma  \\
1+\rho\chi  & 1+\frac{\mu+\nu}{2}\chi  &\frac{\mu}{2}
\end{pmatrix} .
\end{equation*}
Determinant of this  matrix is
\ba\n{det}
\det U =-\sigma [\sigma+2\mu+(\rho\sigma-\sigma\mu+\mu^2+\mu\nu)\chi] .
\ea

At the transition point, when $\chi=0$ and $\dot{\chi}=0$, the system of these equations is
\ba\label{510}
&-\mu q+\sigma v = \Lambda-\rho p-\nu u ,\\
&2q-\sigma \ddot{\chi}=-p ,\\
&q+v+\frac{\mu}{2}\ddot{\chi}=-u .
\ea
At this point $\det U=-\sigma (\sigma+2\mu)$. If $\sigma\neq 0$ and $\sigma+2\mu\neq 0$ relations \eq{510} allow one to find second derivatives $\ddot{a}$, $\ddot{B}$, $\ddot{\chi}$. This means that at least in the vicinity of the transition point there exists a unique regular solution.

\section{Phase I: $v\le \Lambda$ constraint}

\subsection{Transition between sub- and supercritical regimes}

When $\det U\ne 0$ the system  equations  (\ref{LIN_C}), (\ref{E11}) and (\ref{E33}) can be presented in an equivalent form resolved with respect to the second derivatives $\ddot{a}$, $\ddot{B}$ and $\ddot{\chi_{1}}$.
We call set of equations normal.
The initial conditions for this normal set of equations are
\bea
a(\tau_0)&=&a_0, ~~~~\, \dot{a}(\tau_0)=\dot{a}_0,\nonumber\\
B(\tau_0)&=&B_0, ~~~\dot{B}(\tau_0)=\dot{B}_0,     \n{IC} \\
\chi(\tau_0)&=&\chi_0, ~~~~ \dot{\chi}(\tau_0)=\dot{\chi}_0 .\nonumber
\eea
If the functions in the right-hand side of the normal equations are regular in the vicinity of a point (\ref{IC}) then, this set of equations has a regular unique solution.

In order to satisfy condition $\det U\ne 0$ the constant $\sigma$ which enter the linear constraint (\ref{LIN_C}) should be nonzero. The simplest choice which satisfies this property is
\be \n{PPP}
\Phi=  v -\Lambda=0 .
\ee

Let us discuss now a transition between sub- and supercritical regimes for this choice of the constraint function.
In what follows it is very convenient to use dimensional quantities and dimensional form of the equations. It can be done by using the fundamental length scale $\ell$. We denote by hat over a quantity its dimensionless form. Thus we have
\bea
&\hat{p}=\ell^2 p\hh \hat{q}=\ell^2 q\hh \hat{u}=\ell^2 u\hh \hat{v}=\ell^2 v\, ,\nonumber\\
&\hat{\chi}_{1}=\chi_{1}\hh \hat{a}=a/\ell\hh \hat{B}=B\hh  \hat{t}=t/\ell\, ,\n{DIM}\\
&\hat{M}=M/\ell\hh \hat{\Lambda}=\ell^2\Lambda=1 .\nonumber
\eea
In these variables the constraint equation (\ref{TRAN}) takes a simple form
\be\n{vvv}
\hat{v}=1 .
\ee

For our problem a subcritical solution coincides with metric (\ref{a2}). We denote by $\tau_0$ the time when the transition from subcritical to supercritical regime occurs. Then (\ref{PPP}) can be written in the form
\be\n{TRAN}
\hat{v}_0=\hat{p}_0=-2\hat{q}_0=-2\hat{u}_0=\frac{2\hat{M}}{\hat{a}_0^3}=1 .
\ee
Hence
\be\n{aa00}
\hat{a}_0=m\equiv (2\hat{M})^{1/3} .
\ee
In what follows we assume that the transition to the supercritical regime happens inside the horizon of the black hole, so that $m>1$.

Using  relation
\be
\frac{dr}{d\tau}=-\sqrt{f}\hh B(\tau_0)=\sqrt{f(r_0)} ,
\ee
one can write the initial conditions (\ref{IC}) at the transition point as follows
\bea
&\hat{a}_0=m\hh \frac{d\hat{a}}{ d\hat{\tau}}|_0=-\sqrt{m^2-1} ,\nonumber\\
&\hat{B}_0=\sqrt{m^2-1}\hh \frac{d\hat{B}}{ d\hat{\tau}}|_0=\frac{1}{2}m ,\\
&\hat{\chi_{1}}_0=0\hh \frac{d\hat{\chi_{1}}}{d\hat{\tau}}|_0=0 .
\eea

\subsection{Field equations and their solutions}

Starting from now we shall be working only with dimensionless quantities. Since presence of many hats in formulas makes them unwieldy, we adopt the following agreement. We simply omit all the hats in the intermediate results.
One can always easily restore dimensions in the final results by restoring the hats over all the quantities and after this using relations (\ref{DIM}).

Let us write a system of equations for the supercritical regime. The constraint (\ref{vvv}) gives
\be \n{EBB}
\ddot{B}-B=0 .
\ee
Putting  $\rho=\mu=\nu=0$, $\sigma=1$ and using \eq{EBB} one can write (\ref{E33b}) as follows
\be \n{Eaa}
\ddot{a}+\frac{\dot{B}}{B}\dot{a}+a=0 .
\ee
We denote by $\chi_1$ a control function associated with the constraint $v=0$. Then \eq{E00} gives
\be \n{Echi}
\frac{\dot{B}}{ B}\dot{\chi}_{1}+ (2u-1)\chi_{1}=p+2u .
\ee
Expressions for $p$ and $u$ which enter this equation are given by (\ref{pquv}).
The set of three equations (\ref{EBB})-- (\ref{Echi}) with initial conditions
\bea \n{ICV}
&{a}_0=m\hh \dot{a}_0=-\sqrt{m^2-1} ,\nonumber\\
&{B}_0=\sqrt{m^2-1}\hh \dot{B}_0=\frac{1}{2}m\hh\chi_{1,0}=0 .
\eea
The condition $\dot{\chi}_{1,0}=0$ at the transition point follows from (\ref{Echi}).

A solution of (\ref{EBB}) satisfying (\ref{ICV}) is
\be\label{Bcosh}
B(\tau)=\frac{1}{2} \sqrt{3m^2-4} \cosh(\tau-\tau_0+\phi)\, ,
\ee
where
\be
\tanh\phi=\frac{m}{2 \sqrt{m^2-1}} .
\ee
Substituting expression for $B$ into (\ref{Eaa}) one gets
\be \n{Eaa_1}
\ddot{a}+\tanh (\bar{\tau})\dot{a}+a=0
\hh \bar{\tau}=\tau-\tau_0+\phi
 .
\ee
A solution of this equation is
\ba\n{atau0}
a(\tau)=\frac{1}{\sqrt{\cosh(\bar{\tau}})}
\Big[
&C_P P_{-\frac{1}{2}}^{\ii\frac{\sqrt{3}}{2}}\big(\tanh\bar{\tau}\big)\\
+&C_Q Q_{-\frac{1}{2}}^{\ii\frac{\sqrt{3}}{2}}\big(\tanh\bar{\tau}\big)
\Big] .
\ea
Constants $C_P$ and $C_Q$ can be fixed by the initial conditions (\ref{ICV}) and have the form
\bea\nonumber
C_P=\frac{\sqrt{\cosh\phi}}{W\sqrt{\tanh^2\phi-\frac{1}{4}}}\Big[&&Q_{-\frac{1}{2}}^{\ii\frac{\sqrt{3}}{2}}\big(\tanh\phi\big)\\
-(1-i\sqrt{3})\tanh\phi\, &&Q_{\frac{1}{2}}^{\ii\frac{\sqrt{3}}{2}}\big(\tanh\phi\big)\Big],   \nonumber\\
C_Q=-\frac{\sqrt{\cosh\phi}}{W\sqrt{\tanh^2\phi-\frac{1}{4}}}\Big[&&\, P_{-\frac{1}{2}}^{\ii\frac{\sqrt{3}}{2}}\big(\tanh\phi\big) \nonumber\\
-(1-i\sqrt{3})\tanh\phi\, && P_{\frac{1}{2}}^{\ii\frac{\sqrt{3}}{2}}\big(\tanh\phi\big)\Big], \nonumber
\eea
where
\bea
W&=&\frac{1+e^{-\pi\sqrt{3}}}{\pi}\Big[\Gamma\Big(\frac{1}{2}+\ii\frac{\sqrt{3}}{2}\Big)\Big]^2 .
\eea
\begin{figure}[!hbt]
    \centering
      \includegraphics[width=0.40 \textwidth]{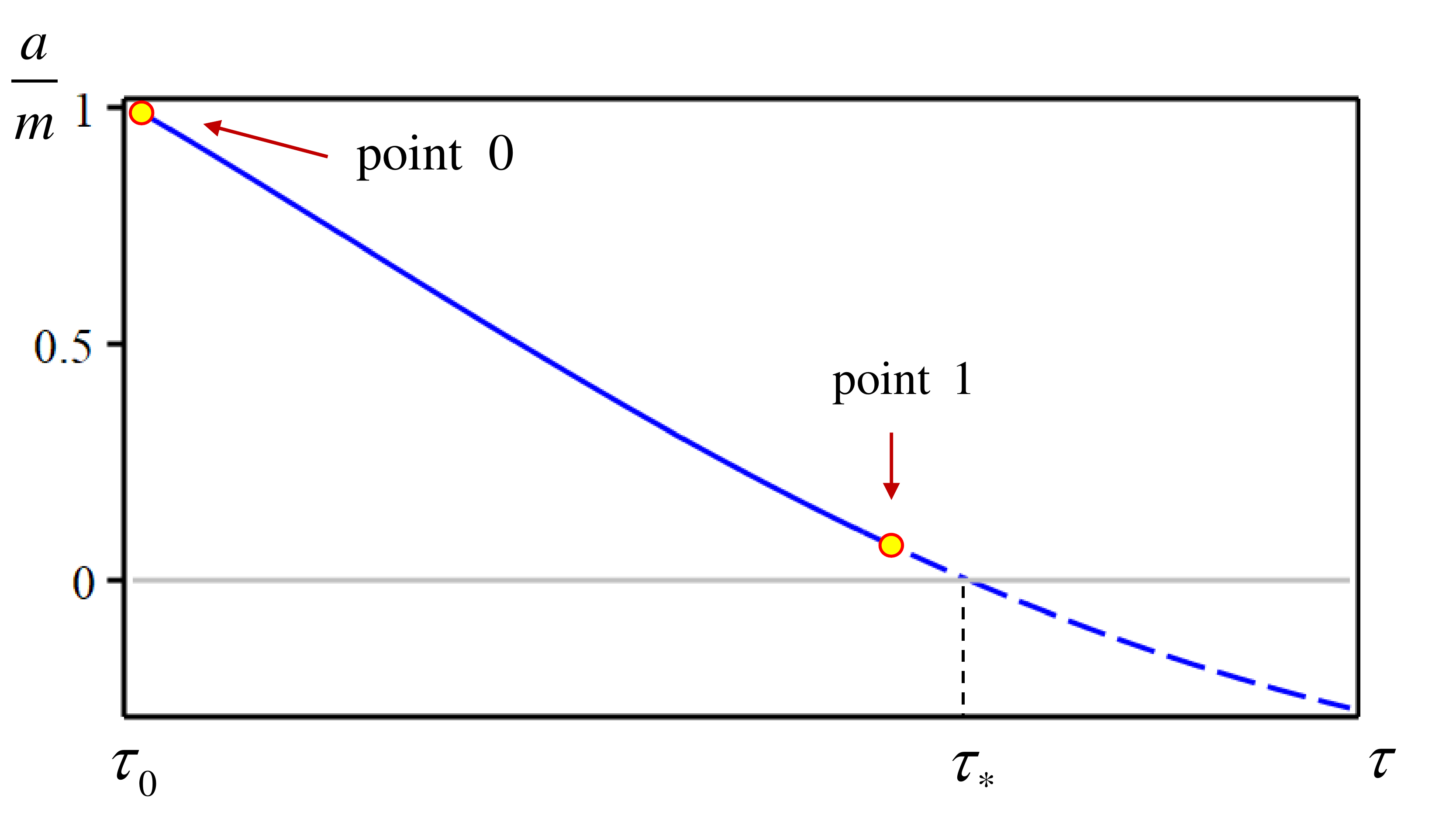}
    \caption{Typical evolution of the scale parameter along the primary constraint. Before the moment when $a(\tau)$ vanish, the secondary constraint enters the play to limit growing curvature invariants. }
    \label{a-tau}
\end{figure}
Note that the solution \eq{atau0} is real despite of the complex parameters entering the expression.
For large $m\gg 1$ the asymptotic of $C_P$ and $C_Q$ coefficients is
\ba
&C_P=(0.40805+1.6113~\ii )\, m +O(m^{-1}),\\
&C_Q=(114.14-23.130~\ii )\, m +O(m^{-1}) .
\ea
Fig.\,\ref{a-tau} shows a plot of $a(\tau)$ computed for $m=1000$. Qualitatively the plots look very similar for all values of $m>2$.  One can see that $a(\tau)$ at this phase is a monotonically decreasing function and there always exists such a moment of time $\tau=\tau_*$ when it vanishes. In the limit $\tau\to \tau_*$ invariants $p$ and $q$ infinitely grow. In order to prevent this, in what follows we shall impose an additional constraint. We discuss a choice of this constraint in the next section. Here we just continue our study of the stage with one constraint $v=1$.
\begin{figure}[!hbt]
    \centering
\includegraphics[width=0.4 \textwidth]{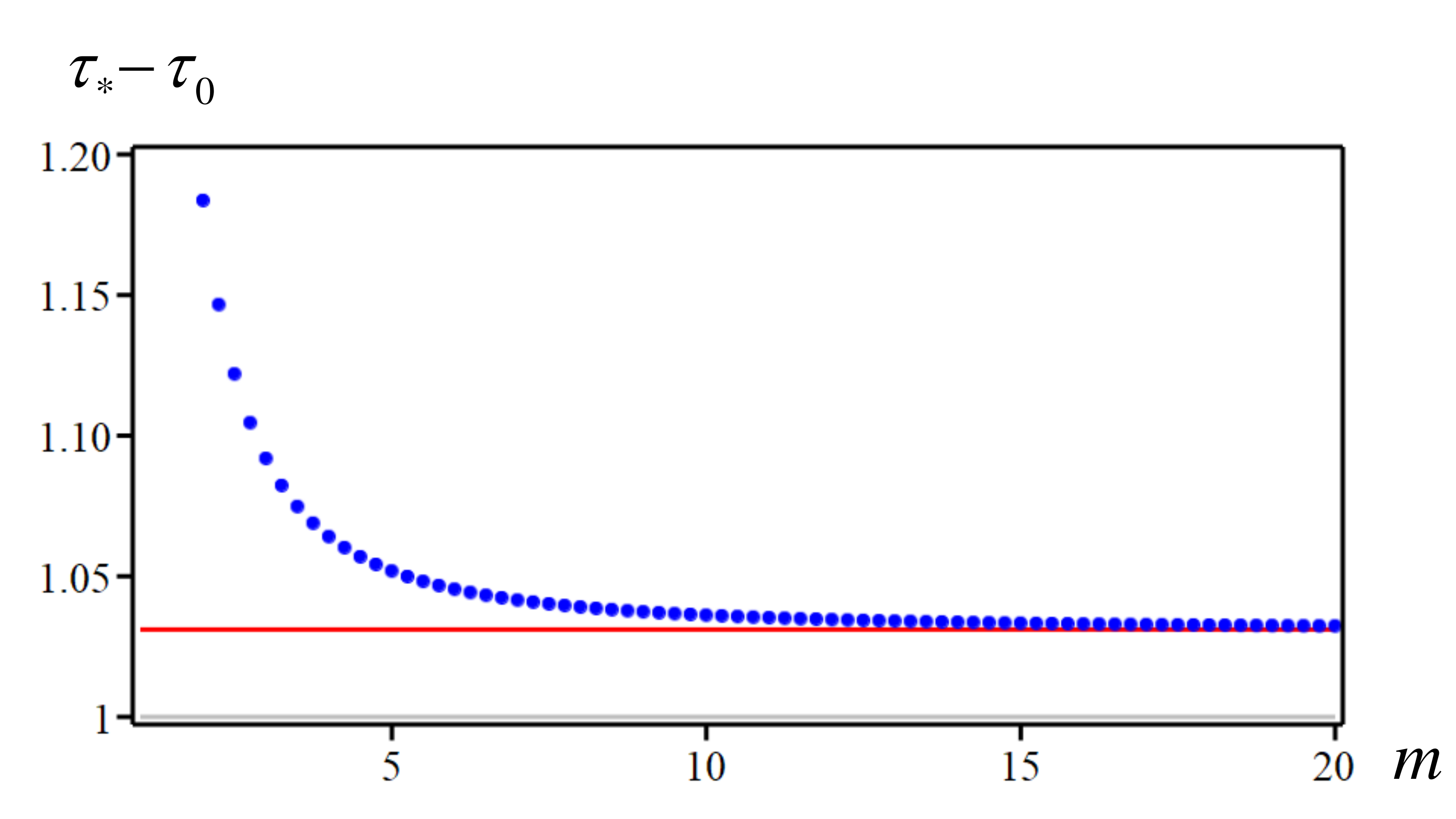}
    \caption{The dependence of $\tau_*$ on the parameter $m$. Its asymptotic value at large $m$ is $\tau_*=\tau_0+1.031$.}
    \label{tau-star}
\end{figure}

The control function $\chi_{1}(\tau)$ can be found by solving equation (\ref{Echi}). A solution satisfying the initial condition $\chi_{1}(\tau_0)=0$ is
\be\label{chi1}
\chi_{1}(\tau)=1+\frac{\dot{B}}{a^2} \left[
-\frac{a^2_0}{\dot{B}_0} +\int_{\tau_0}d\tau  \frac{a^2 B}{\dot{B}^2}(1+p)
\right] .
\ee
Note that due to \eq{ICV}
\ba
\frac{a^2_0}{\dot{B}_0}=2m .
\ea

\subsection{Case of large $m$}

When the radius $r_0$ where the Schwarzschild metric reaches the critical curvature is much larger than the critical length $\ell$ the dimensionless parameter $m$, (\ref{aa00}), is large. In this regime $\sqrt{m^2-1}=m+O(m^{-1})$. Neglecting terms $O(m^{-1})$ in (\ref{ICV}) one gets the following initial conditions
\be \n{ICVm}
{a}_0=-\dot{a}_0=B_0=2\dot{B}_0=m\, ,
\ee
Let us denote
\be
\bar{a}=\frac{a}{m}\hh \bar{B}=\frac{B}{m} \, ,
\ee
then one has
\ba \n{DIMLESS}
&& \ddot{\bar{B}}-\bar{B}=0\hh \ddot{\bar{a}}+\frac{\dot{\bar{B}}}{ \bar{B}}\bar{a}+\bar{a}=0\, ,\\
&&\frac{\dot{\bar{B}}}{\bar{B}}\dot{\chi}_{1}+(2u-1)\chi_{1}=p+2u\, ,\\
&&{\bar{a}}_0=-\dot{\bar{a}}_0=\bar{B}_0=2\dot{\bar{B}}_0=1 .
\ea
In these equations
\be
u=\frac{\dot{\bar{a}}}{\bar{a}}\frac{\dot{\bar{B}}}{ \bar{B}} .
\ee
Since $m$ is large, one can use the following approximate expression for $p=(\dot{\bar{a}}/\bar{a})^2$.
Thus when $m$ is large both field equations and initial conditions in (\ref{DIMLESS}) do not depend on $m$ and are universal in this sense.

Solutions for $\bar{B}$ and $\bar{a}$ are
\ba\n{atau}
\bar{B}(\tau)=&\frac{\sqrt{3}}{2}\cosh(\bar{\tau}) \hhh
\tanh\phi=\frac{1}{2} \hhh
\bar{\tau}=\tau-\tau_0+\phi ,\\
\bar{a}(\tau)=&\frac{\left[
c_P P_{-\frac{1}{2}}^{\ii\frac{\sqrt{3}}{2}}(\tanh\bar{\tau})+
c_Q Q_{-\frac{1}{2}}^{\ii\frac{\sqrt{3}}{2}}(\tanh\bar{\tau})
\right]}{\sqrt{\cosh\bar{\tau}}} .
\ea
Here
\be
c_P=0.40805+1.6113~\ii \hhh c_Q = 114.14-23.13~\ii .
\ee

\begin{figure}[!hbt]
    \centering
      \includegraphics[width=0.4 \textwidth]{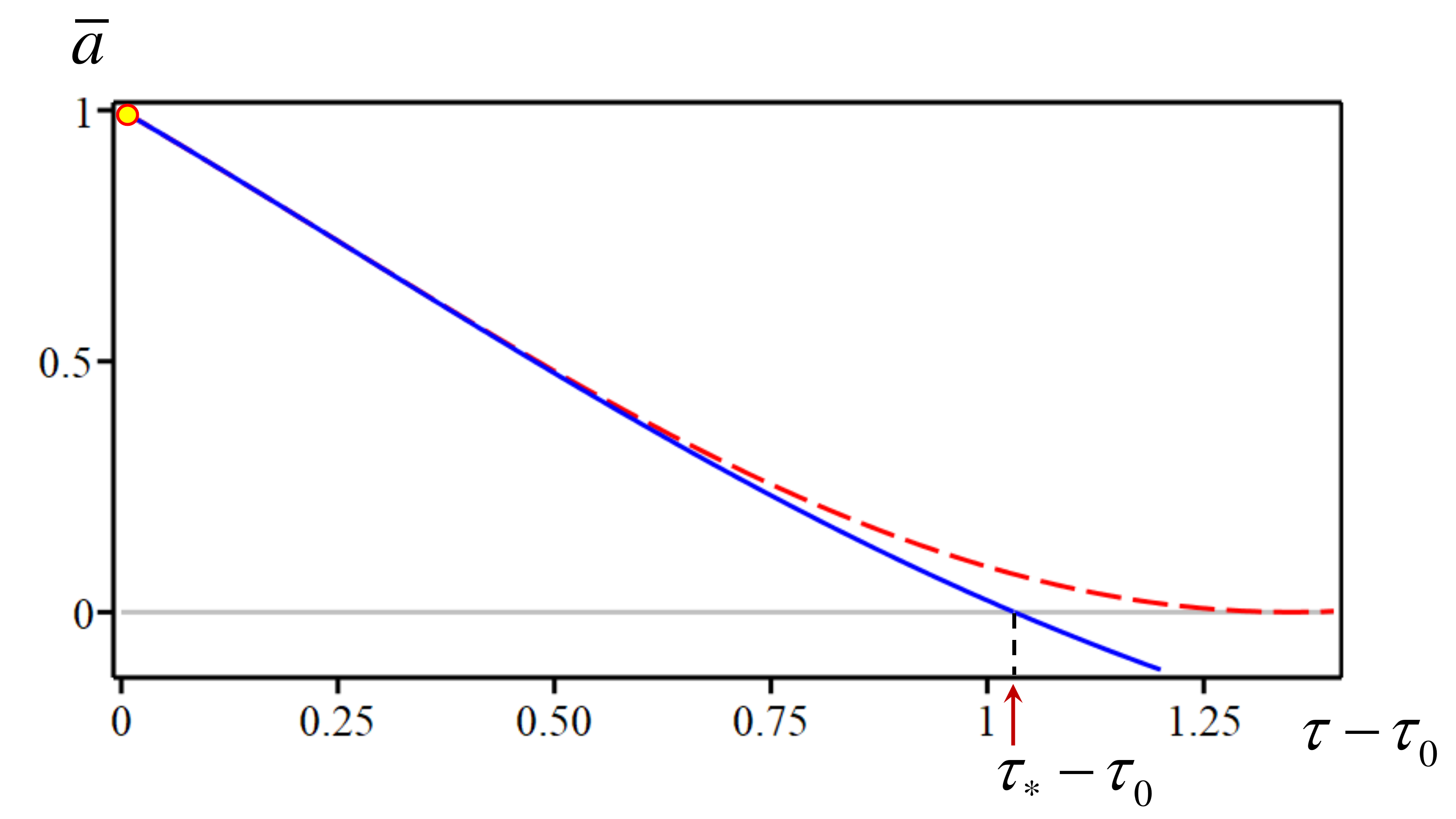}
    \caption{The exact solution $\bar{a}(\tau)$ (solid line) and the approximate one (dashed line). The scale factor vanishes at the moment $\tau_*$. But before this moment the secondary constraint changes the dynamics and stops the collapse of $\bar{a}(\tau)$.}
    \label{Approximation_a}
\end{figure}

A plot of $\bar{a}(\tau)$  is shown by the solid line in Fig.\,\ref{Approximation_a}. The function $\bar{a}(\tau)$ monotonically decreases starting from the initial value $\bar{a}_0=1$  and becomes zero at $\tau=\tau_*$, where $\tau_*-\tau_0=1.03$.
 The invariant $q$ evaluated on the exact solution starts from $q(\tau_0)=-1/2$ and then grows with time. It vanishes at  $\tau\approx \tau_0+0.26966$ and becomes positive for later times.  In the vicinity of $\tau_*$ both invariants $p$ and $q$ infinitely grow.

The approximate solution, corresponding to the linear asymptotic of $\bar{a}$ at $\tau=\tau_0$
is given by the dashed line in Fig.\,\ref{Approximation_a}. It has a simple form
\ba\n{aaaaa}
\bar{a}=\frac{3}{4}\Big[1-\sin\Big(\sqrt{2}(\tau-\tau_0)-\arcsin\frac{1}{3}\Big)\Big],
\ea
which approximates the exact solution $\bar{a}(\tau)$  \eq{atau} in the range $\tau\in(\tau_0,\tau_0+0.3)$ with the accuracy $10^{-3}$.
The basic curvature invariants $p$ and $q$ calculated for this solution obey  the equation
\be \n{pqAP}
p=2(1+q) ,
\ee

\begin{figure}[!hbt]
    \centering
      \includegraphics[width=0.4 \textwidth]{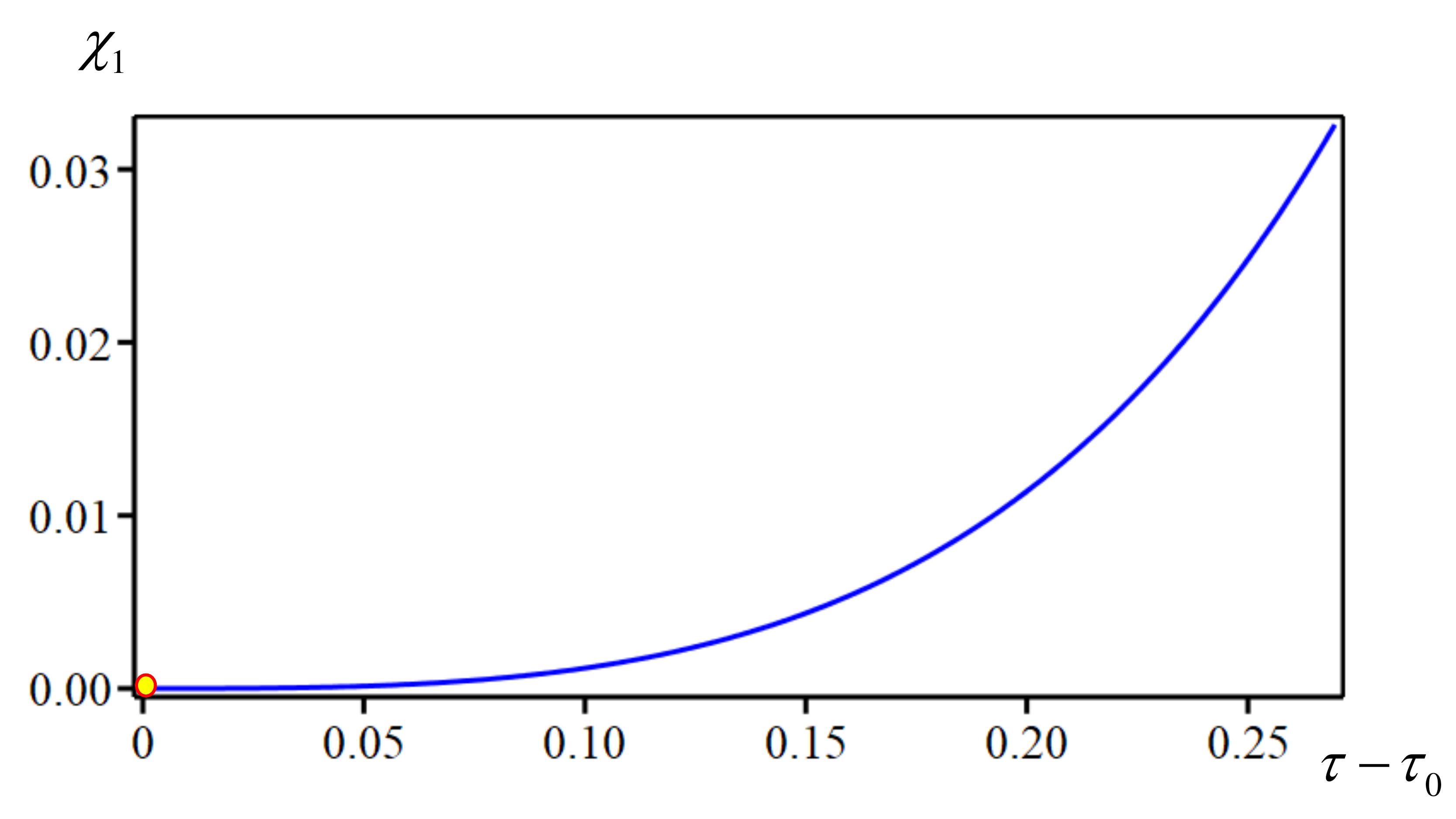}
    \caption{The limit of large $m$ for the control parameter $\chi_1(\tau)$. We draw the plot in the interval $0\le \tau-\tau_0\le 0.27$. For larger $\tau$ it continues to grow monotonically.}
    \label{Figchi1}
\end{figure}

The solution \eq{chi1} for the control function $\chi_1$ takes the form
\be\label{chi1a}
\chi_{1}(\tau)=1+\frac{\dot{\bar{B}}}{\bar{a}^2} \left[
-2 +\int_{\tau_0}^{\tau}d\tau  \frac{\bar{B}}{\dot{\bar{B}}^2}(\dot{\bar{a}}^2+\bar{a}^2)
\right] .
\ee
At the initial point both the control parameter and its derivative vanish $\chi_{1}(\tau_0)=0$, $\dot{\chi}_{1}(\tau_0)=0$.
Then it monotonically grows with time. This property can be easily seen because for $\tau>\tau_0$ the functions $\bar{B}$, $\dot{\bar{B}}$, and the integrand in \eq{chi1a} are positive definite functions. The behavior of the control parameter  is depicted in Fig.\,\ref{Figchi1}. It means that the control function $\chi_1$ will never vanishes again and the solution for the metric will evolve along this constraint till the moment, when the conditions for the secondary constraint are fulfilled.

\section{Phase II}

\subsection{Field equations}

As we saw in the previous section the constraint $v=1$ does not prevent growth of curvature invariants $p$ and $q$. To restrict them one needs, besides this primary constraint $v=1$, to impose an additional secondary constraint.
Namely, we assume that the first supercritical phase ends at some moment of time $\tau=\tau_1$. After this moment the supercritical solution evolves preserving both, primary and secondary constraints. At the second phase there exist two control functions $\chi_1$ and $\chi_2$ associated with both constraints.
Their evolution is determined by gravitational equations.
The initial value of the primary control function $\chi_1$ at $\tau=\tau_1$ can be found by using solution \eq{chi1}, while control function $\chi_2$ and its derivative vanish at this point.
	
\subsubsection{Primary constraint}

This constraint $v=1$ is the same as at the first phase
\be \n{eqBB2}
\ddot{B}-B=0 .
\ee
Its solution, obtained for phase~I
\ba\n{BB2}
&B(\tau)=\frac{1}{2} \sqrt{3m^2-4} \cosh(\tau-\tau_0+\phi)\, ,\\
&\tanh\phi=\frac{1}{2}\frac{m}{\sqrt{m^2-1}}\, ,
\ea
evidently satisfies the required continuity conditions at $\tau=\tau_1$.

\subsubsection{Secondary constraint}

For the secondary constraint at phase~II we use a constraint function which is linear in the curvature invariants. We choose it in the form
\be \n{pql}
p-\mu q=\lambda .
\ee
Here, as earlier, we use dimensionless quantities normalized by the length scale $\ell$,
$\mu\in (0,1)$ is a dimensionless constant and $\lambda=\Lambda_2/\Lambda_1$. Note that at the second phase $p>1$ and $q>-{1}/{2}$. Therefore
\ba\label{lam}
\lambda>1+\frac{\mu}{2} .
\ea
Equation (\ref{pql})  gives the following equation
\be\n{aaaa}
\frac{\dot{a}^2}{ a^2}+\frac{1}{a^2}-\mu \frac{\ddot{a}}{a}=\lambda .
\ee
Solution for $a(\tau)$ in the phase~I and condition of continuity of $a(\tau)$ and  $\dot{a}(\tau)$ at the second transition point $\tau=\tau_1$ uniquely specifies the metric function $a(\tau)$ during the second phase.

\subsubsection{$(\theta,\theta)$ gravitational equation}

At phase II $(\theta,\theta)$ equation has the form
\bea\n{tteq}
\frac{\ddot{a}}{a}&+&\frac{\dot{a}\dot{B}}{aB}+1\nonumber\\
&=&-\Big[\frac{\dot{a}}{a}\dot{\chi}_2+\Big(
\frac{\ddot{a}}{a}+\frac{\dot{a}\dot{B}}{aB}
+\frac{\dot{a}^2}{a^2}+\frac{1}{a^2}\Big)\chi_{2}     \Big]\\
&&- \frac{\mu}{2}\Big[ \ddot{\chi}_2
+\Big(2\frac{\dot{a}}{a}+2\frac{\dot{B}}{B}\Big)\dot{\chi}_2
+\Big(
2\frac{\dot{a}\dot{B}}{aB}+1\Big)\chi_{2}
\Big] .\nonumber
\eea
For known metric functions $a(\tau)$ and $B(\tau)$ this is a second-order linear ordinary differential equation for $\chi_2(\tau)$. Two integration constants in its solution are fixed by the initial conditions
\be
\chi_2(\tau_1)=\dot{\chi}_2(\tau_1)=0 .
\ee

\subsubsection{$(\tau,\tau)$ gravitational equation}

$(\tau,\tau)$  gravitational equation is of the form
\bea\label{EE00}
\frac{\dot{a}^2}{a^2}&&+\frac{1}{a^2}+2\frac{\dot{a}\dot{B}}{aB}
=\Big(-\frac{\dot{a}^2}{a^2}+\frac{1}{a^2}
\Big)\chi_{2} \nonumber\\
&&-\mu\Big[
\frac{\dot{a}}{a}\dot{\chi}_2+\Big(\frac{\dot{a}^2}{a^2}
+\frac{\dot{a}\dot{B}}{aB}\Big)\chi_{2}\Big]-\lambda\chi_{2}\\
&&+\frac{\dot{B}}{B}\Big[\dot{\chi}_{1}+2\frac{\dot{a}}{a}\chi_{1} \Big]-\chi_{1} .
\nonumber
\eea
For known functions $a(\tau)$, $B(\tau)$ and $\chi_{2}(\tau)$ this is a first-order linear differential equation for $\chi_{1}(\tau)$. The expression \eq{chi1} at $\tau_1$ determines the initial conditions for $\chi_{1}(\tau_1)$.
Hence,  the control function $\chi_{1}(\tau)$ is uniquely defined at the second phase.

Let us summarize: A set of four equations (\ref{eqBB2}), (\ref{aaaa}), (\ref{tteq}) and (\ref{EE00}) with the above described  initial conditions uniquely determines four functions $a(\tau)$, $B(\tau)$, $\chi_1(\tau)$ and $\chi_2(\tau)$ at the phase~II. Let us discuss now properties of this solution.

\section{Phase~II: Solution of the secondary constraint}

\begin{figure}[!hbt]
    \centering
      \includegraphics[width=0.30 \textwidth]{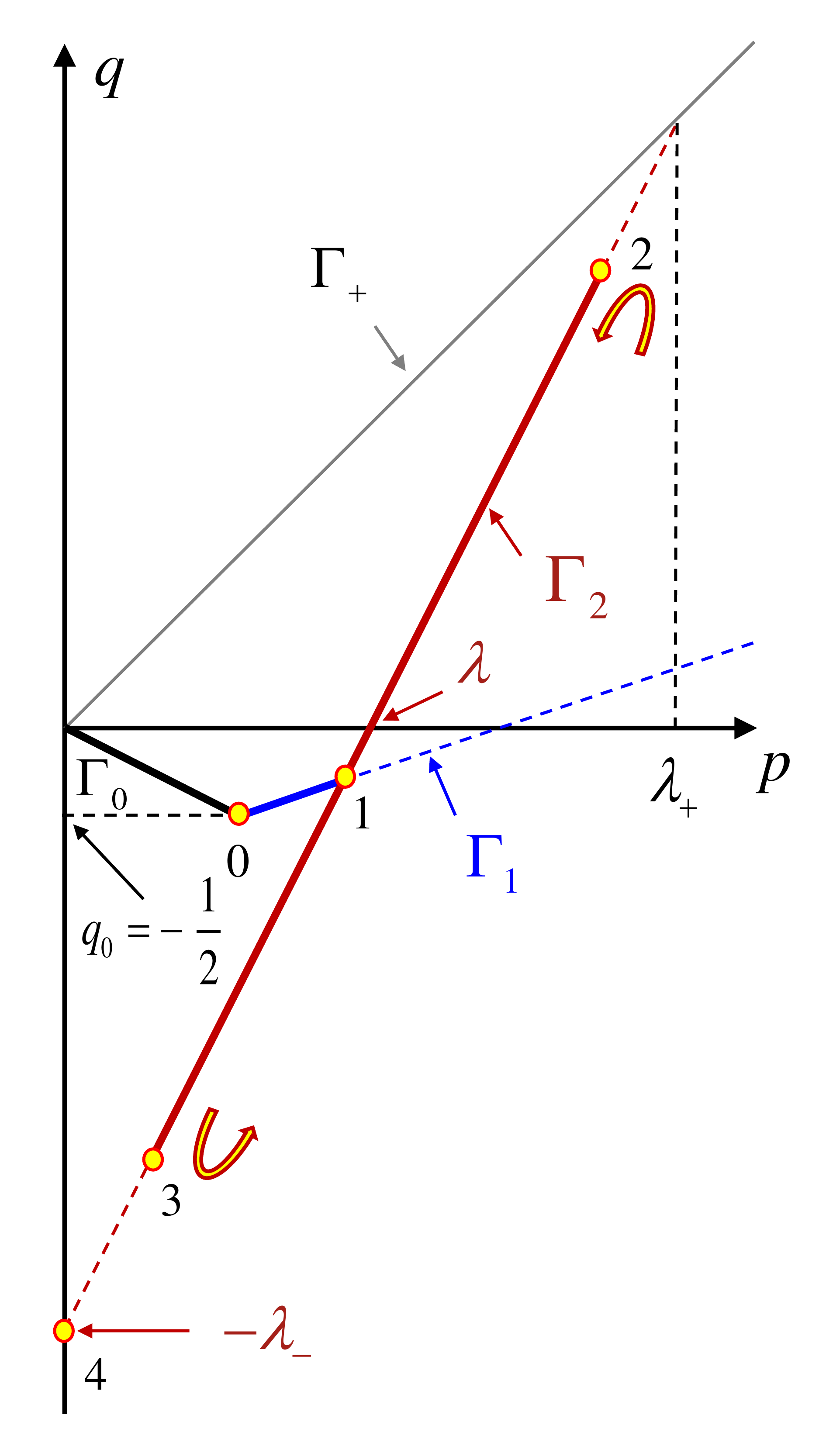}
    \caption{At the beginning the system evolves along $\Gamma_0$ governed by the pure Einstein equations. At the point $0$ the conditions for the primary constraint are satisfied and further evolution goes along the line $\Gamma_1$ till the point $1$, where the secondary constraint enters the play. After this point the system evolves along the line  $\Gamma_2$ towards the point $2$. This is the bouncing point, corresponding to a minimal scale parameter $a(\tau)$. After this bounce the system oscillates along $\Gamma_2$ between the  bouncing points $2$ and $3$.}
    \label{Two_Stages}
\end{figure}

The solution of the primary constraint was already described (see (\ref{BB2})). In this section we discuss solutions of the secondary constraint (\ref{pql}) at phase~II. For this purpose it is convenient to use a representation of such  solutions on two-dimensional $(p,q)$ plane shown in Fig.~\ref{Two_Stages}.

This figure contains a set of lines connected with the evolution of the system. A straight line $\Gamma_0$ represents  a subcritical Schwarzschild solution. On this line $q=-\frac{1}{2}p$. At a point $0$ where $p=1$ and $q=-1/2$ the invariant $v$ reaches its critical value $v=1$. It happens at point $0$ where the first phase of the supercritical solution starts. Line $\Gamma_1$ shows $p$ and $q$ during this phase. The second phase starts at point $1$. At this phase the supercritical solution obeys both constraints $v=1$ and $p-\mu q=\lambda$ and $\Gamma_2$ is a straight line representing this solution. This line intersects $p$ axis at $p=\lambda$. For $0<\mu<1$ the line $\Gamma_1$,  being continued to larger value of $p$,  crosses the line $\Gamma_+$, where $p=q$, at the point
\be
p=q=\lambda_+=\frac{\lambda}{1-\mu} .
\ee
Similarly, if one continues $\Gamma_2$ to a small value of $p$, then for $\mu>0$
it  intersects $q$-axis at $q=-\lambda_-=-\lambda/\mu<0$.
In what follows we impose the following restriction on the parameter $\mu$, $0<\mu<1$.

\subsection{Solving secondary constraint equation}

\subsubsection{Function $a(\tau)$ at the beginning of the second phase}

To estimate the function $a(\tau)$ at the second transition point 1 for large $m$ one can use the approximate equation (\ref{pqAP}). In this approximation point 1 is the intersection of two straight lines $\Gamma_1$ and $\Gamma_2$. Simultaneous solution of two equations $p=2(1+q)$ and $p-\mu q=\lambda$ gives
\be\n{qq1}
q_1=-{2-\lambda\over 2-\mu}\, .
\ee
Using the approximate solution (\ref{aaaaa}) one finds
\be \n{aa1}
\bar{a}_1={3\over 2}\ {1\over 2+q_1}\, .
\ee
Relations (\ref{qq1}) and (\ref{aa1}) give
\be \n{aamm}
a_1=m\bar{a}_1\hh \bar{a}_1={3\over 2}\ {2-\mu\over 2+\lambda-2\mu}\, .
\ee
It is easy to check that $a_1\le a_0$ for $\lambda\ge 1+\mu/2$.
Relation (\ref{aamm}) shows that for the parameters $\mu$ and $\lambda$ of order of one, the initial value of function $a(\tau)$ at the beginning of the second phase $a_1$ is large (proportional to $m$).

\subsubsection{Solution}

To solve equation (\ref{pql}) we denote
\be
p=p(y)\hh y=\ln(a^2/a_1^2)\, ,
\ee
where $a_1=a(\tau_1)$
Then using definitions of $p$ and $q$ one gets the following equation
\be \n{ppqq}
\frac{dp}{dy}=q-p .
\ee
We choose the second transition point $1$ to lie below $\Gamma_+$, so that $\frac{dp}{dy}<0$ at this point. We also assume that $\dot{a}_1<0$, then in the vicinity of $1$ the function $a$ decreases and $p$ increases. This means that a point representing the solution on $(p,q)$ plane moves up along the line $\Gamma_2$.
Using (\ref{ppqq}) one gets
\be \n{yypp}
y=\int_{p_1}^p \frac{dp}{q-p} .
\ee
Here $p_1$ is a value of $p$ at point $1$.
This relation gives
\be
\frac{\lambda_{+} -p}{ \lambda_+ -p_1}=\left(\frac{a}{a_1}\right)^{\gamma}
\hh \gamma=\frac{2(1-\mu)}{\mu}   .
\ee
One gets
\be\label{plam}
p=\lambda_+ -(\lambda_+ -p_1)\left(\frac{a}{a_1}\right)^{\gamma} .
\ee

Using the definition of $p$ \eq{pquv} one has
\be\n{eqap}
\dot{a}^2=a^2 p-1 .
\ee
Thus $a^2p\ge 1$ and
\be
p=\frac{1}{a^2}
\ee
at the turning points where $\dot{a}=0$. It is easy to check that there exist two turning points corresponding to minimum and maximum values of $a$. At the point of the minimum one has
\be\n{aami}
a_{2}\approx \frac{1}{\sqrt{\lambda_+}}\Big[1+\frac{1}{2}\Big( 1-\frac{p_1}{\lambda_+}\Big) \big(\sqrt{\lambda_+} a_1\big)^{-\gamma}\Big] .
\ee
This expression is written in an approximation when the quantity $ \lambda_+ a_1^2$ is large.
The maximum value of $a$ is
\be\n{aama}
a_{3}\approx A a_1 \hh A=\Big( 1-\frac{p_1}{\lambda_+}\Big)^{-\frac{1}{\gamma}} .
\ee

Let us summarize. A point in $(p,q)$ plane representing a supercritical solution $a(\tau)$ at phase~II moves up along $\Gamma_2$. The value of $a$ decreases until it reaches its minimal value at point $2$ (see Fig.\,\ref{Two_Stages}). We denote by $\tau_2$ time when it happens. Relation (\ref{yypp}) shows that this point always is below the line $\Gamma_+$. For large $a_1$, that is when $a_1\gg \lambda_+^{-1/2}$, point $2$ is located very close to $\Gamma_+$. Later for $\tau>\tau_2$ the point representing the solution moves down along $\Gamma_2$ to smaller value of $p$. In a general case, the control function $\chi_2(\tau)$ cannot become zero simultaneously with its derivative, so that this supercritical solution $a(\tau)$ always remains on the line $\Gamma_2$. It passes through point $1$ and reaches its maximum  value $a_{\max}$ at point $3$. This happens at time $\tau_3$. After this, the function $a(\tau)$ decreases again. This motion is periodic and its period is
\be
T=2(\tau_3-\tau_2)=2\int_{a_{2}}^{a_{3}} \frac{\dd a}{\sqrt{p(a)a^2-1}} .
\ee
It should be emphasised that in this approach we put constraints only on the curvatures $p$, $q$, and $v$. However, the curvature invariant $u=\frac{\dot{a}\dot{B}}{a B}$ is finite automatically. This is because primary constraint guarantees that $|\dot{B}/B|\le 1$, while the finite value of $p$ and the property, that scale factor $a$ never vanishes on the solution in question, makes $|\dot{a}/a|\le$ finite too.

\subsection{Special case $\mu=1/2$}

There is a special choice of the parameter $\mu=1/2$, for which equation (\ref{plam})-(\ref{eqap})  can be integrated analytically. For this choice one also has
\be
\lambda_-=\lambda_+=2\lambda
\hh \gamma=2 .
\ee
This relation implies, that during periodic change of the function $a(\tau)$, the curvature invariants remain in the following intervals
\be
p\in (0,2\lambda)\hh q\in (-2\lambda,2\lambda) .
\ee
Let us discuss this case in more detail.

Let us denote
\be
a=a_1\alpha\, ,
\ee
then equation \eq{eqap} can be written in the form
\ba\n{alp}
&\dot{\alpha}^2=(\lambda_+-p_1)(\alpha^2-\alpha_{2}^2)(\alpha_{3}^2-\alpha^2) ,\\
&\alpha_{2,3}^2=\frac{\lambda_+}{2(\lambda_+-p_1)}\left[
1\mp \sqrt{1-\frac{4(\lambda_+-p_1)}{\lambda_+^2 a_1^2}}
\right]  .
\ea
Here $\alpha_{2}$ and $\alpha_{3}$ correspond to signs minus and plus in \eq{alp} respectively.
Equation \eq{alp} gives
\ba
&\tau-\tau_1=\frac{a_1}{\sqrt{\lambda_+ -p_1}}N(\alpha)\, ,\\
& N(\alpha)=
\int_1^{\alpha} \frac{d\alpha}{\sqrt{(\alpha^2-\alpha_{2}^2)(\alpha_{3}^2-\alpha^2)}} .
\ea
This integral can be calculated exactly with the following result
\be
N(\alpha)=\frac{\ii}{\alpha_-}\left[
F( \phi,k)-F(\phi_1,k)
\right] ,
\ee
where
\ba
\sin\phi= \frac{\alpha}{\alpha_{3}}\hh \sin\phi_1= \frac{1}{\alpha_{3}}\hh k=\frac{\alpha_{3}}{ \alpha_{2}} .
\ea
Here $F(\phi,k)$ is an elliptic integral. Note that for $\alpha_{2}\le\alpha\le\alpha_{3}$ the function $N(\alpha)$ is a real.
The period of oscillations is given by the integral
\ba
 T&=2\frac{a_1}{ \sqrt{\lambda_+ -p_1}}\int_{\alpha_{2}}^{\alpha_{3}} \frac{d\alpha }{ \sqrt{(\alpha^2-\alpha_{2}^2)(\alpha_{3}^2-\alpha^2)}}\\
&= \frac{2 a_1}{ \sqrt{\lambda_+ -p_1}\,\alpha_{3}}K\left( \sqrt{1-\frac{\alpha_{2}^2}{\alpha_{3}^2}}\right)
 .
\ea
Here $K$ is the complete elliptic integral.

\subsection{Case of small $\mu$}

The secondary constraint takes simpler form $p=\lambda$ for $\mu=0$. Let us discuss properties of phase~II supercritical solution for small values of the parameter $\mu$. We demonstrate that  in fact the limit $\mu\to 0$ is quite singular.

For small $\mu$ line $\Gamma_2$ representing a solution on $(p,q)$ plane (see Fig.\,\ref{Two_Stages}) becomes almost vertical. To mark points on this line it is more convenient to use parameter $q$ instead of $p$. This can be done by using the following relation
\be
\lambda_+ -p =\mu(\lambda_+ -q) .
\ee
Equation (\ref{eqap}) can be written in the form
\ba\n{eqap1}
&\dot{\alpha}^2=U(\alpha) , \\
&U(\alpha)=\lambda_+\alpha^2-(\lambda_+- q_1)\mu \alpha^{\frac{2}{\mu}}
-\frac{1}{a_1^2} .
\ea
Here $\alpha=a/a_1$ and $a_1$ is the value of $a(\tau)$ at the beginning of phase two $\tau=\tau_1$, while $q_1$ is the value of $q$ at this time.

\begin{figure}[!hbt]
    \centering
      \includegraphics[width=0.4 \textwidth]{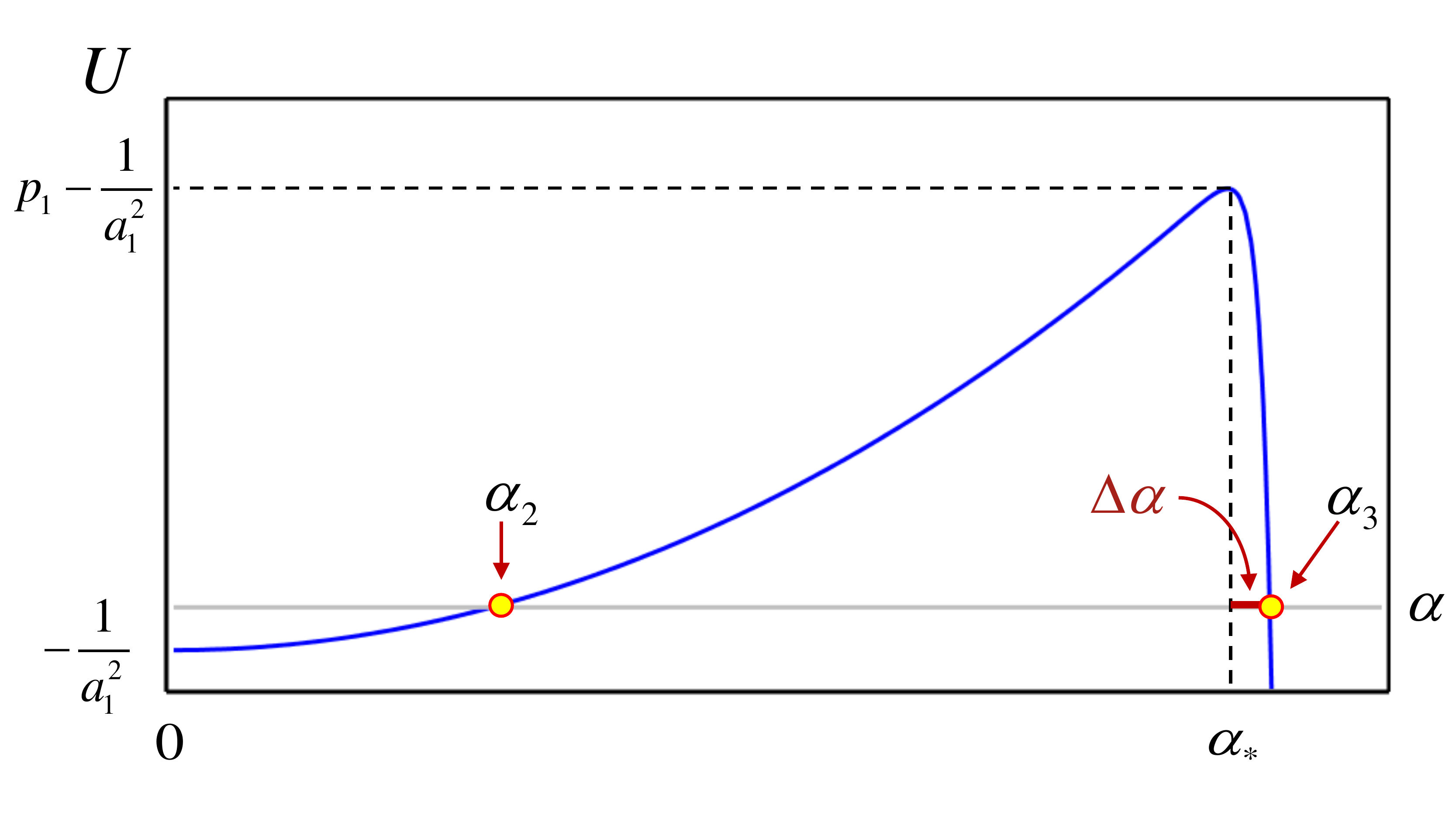}
    \caption{The potential function $U(\alpha)$ for small $\mu$.}
    \label{Small_mu}
\end{figure}

Fig.\,\ref{Small_mu} shows a plot of function $U(\alpha)$ for small $\mu$. The potential function $U(\alpha)$ has a maximum when $dU/d\alpha=0$. It happens at
\be
\alpha_*=\Big[1-\frac{(1-\mu)q_1}{\lambda}\Big]^{-\frac{\mu}{2}}\approx 1+\frac{\mu q_1}{2\lambda},
\ee
which is for small $\mu$ is very close to unity. The values $\alpha_2$ and $\alpha_3$ where $U(\alpha_{2,3})=0$ correspond to the turning points, where $\alpha$ reaches its minimum and maximum values, respectively. For  $a_1\gg 1$ one can use relations (\ref{aami}) and (\ref{aama}) to find the values of $\alpha_2$ and $\alpha_3$. Parameter $A$ which enters the latter expression is slightly larger that $1$.

The function $\alpha(\tau)$ is periodic with the period
\ba
T=2\int_{\alpha_2}^{\alpha_3}\dd \alpha \frac{1}{\sqrt{U(\alpha)}}.
\ea
For small $\mu$ this period in dimensionless units is
\ba
T\approx \frac{2}{\sqrt{\lambda}}\ln (2\sqrt{\lambda}\,a_1) .
\ea
Here we used the property that $\lambda_+=\frac{\lambda}{1-\mu}\approx \lambda$.

At the turning point where $\alpha=\alpha_{3}$ one has
\be
p_{3}=\frac{1}{\alpha_{3}^2 a_1^2} .
\ee
Since $ \alpha_{3}\approx 1$ for large $a_1$,  one has
\be
p_{3}\approx \frac{1}{a_1^2}\ll 1 .
\ee
The invariant $q$ at this point can be calculated either using  relation
\be\n{UUU}
q=\frac{1}{2 \alpha}\frac{dU}{d\alpha}\, ,
\ee
or the secondary constraint equation which gives
\be
q=\frac{p-\lambda}{\mu} .
\ee
One has
\be \n{qqmu}
q=\lambda_+-(\lambda_+-q_1) \alpha^{\frac{2}{\mu}} .
\ee
At $\tau=\tau_1$ we have $\alpha=1$ and $q=q_1$, as expected. For $\tau>\tau_1$ the metric function $\alpha$ decreases and the second term in the right-hand side of (\ref{qqmu}) becomes small very fast. At this stage $q\approx \lambda_+$. After bounce the function $a(\tau)$ increases until it reaches its turning point where it has its maximal value $\alpha_{3}$.
Near this turning point  one has
\be
q_{3}\approx -{\lambda_-}=-\frac{\lambda}{\mu} .
\ee
Thus for large $a_1$ and small $\mu$ the point $(p_{3},q_{3})$ is close to the point $4$ in Fig.\,\ref{Two_Stages}. In the limit $\mu\to 0$ the invariant $q$ grows infinitely near the turning point 3. This case does not satisfy the adopted limiting curvature condition.

\section{Phase~II: Control functions}

The dynamics of the control functions at phase II is described by the gravity equations \eq{tteq} and \eq{EE00} restricted by the constraints. The control function  $\chi_2$ is defined by \eq{tteq} with the initial conditions
\ba
\chi_2(\tau_1)=0 \hh \dot{\chi}_2(\tau_1)=0 .
\ea
The equation \eq{tteq} is the second order ordinary differential equation, therefore its solution is uniquely defined by these initial conditions.
The function $B(\tau)$ is defined by the solution of the primary constraint \eq{EBB} and is given by \eq{Bcosh}. The secondary constraint, in its turn,  defines the function $a{\tau}$. Therefore we unambiguously obtain functions
$B$, $a$, and $\chi_2$. Then their substitution into \eq{EE00} leads to the first order differential equation for the control function $\chi_1$. The initial condition for this function comes from its value at the end of the Phase I
$
\chi_1(\tau=\tau_1).
$
Thus the whole evolution of the system is completely fixed.

If the control parameter $\chi_2$ and its derivative $\dot{\chi}_2$ vanish simultaneously at some point, then the system would slip to the phase I again. However, it is virtually improbable, because the equations for the control parameters explicitly depend on the function $B(\tau)$ which grows irreversibly. So, even if $\chi_2$ vanishes at some moment, in a generic case it's derivative does not vanish and the system will continue to evolve along the secondary constraint.
It means that in the generic case the system will indefinitely stay on the secondary constraint and it permanently oscillates between the points $2$ and $3$ (see Fig.\,\ref{Two_Stages}).

\section{Summary and discussions}\label{SecX}

In this paper we discuss properties of black holes in the limiting curvature theory of gravity \cite{frolov2021bouncing,Frolov:2021kcv}. Namely, we consider four-dimensional spherically symmetric black holes and study properties of their interior in a model, where the curvature invariants are restricted. To satisfy the limiting curvature condition, we modified the Einstein-Hilbert action by adding terms which impose inequality constraints on the curvature invariants. In a general case each of such constraints $\Phi_i\le 0$ is accompanied by a pair of the Lagrange multipliers $\chi_i$ and $\zeta_i$. A solution of the field equations, which are derived from such an extended action, can have
different regimes. In a subcritical regime, where all constraint functions obey conditions $\Phi_i< 0$,
the control functions $\chi_i$ vanish and the equations coincide with the unmodified Einstein equations, while
Lagrange multipliers $\zeta_i$ can be expressed in terms of solutions of these equations. When at least one of the constrains is saturated, the regime is changed. A corresponding control function(s) $\chi_i$ becomes non-zero, while $\zeta_i=0$. Such a regime is called supercritical.

We assumed that a transition from sub- to supercritical regimes happens inside the event horizon of the black hole at some radius $r_0$  where the spacetime curvature reaches its critical value $\Lambda=1/\ell^2$. For $r>r_0$ the solution coincides with Schwarzschild metric and
 $r_0=\ell m=\ell (2M/\ell)^{1/3}$, where $M$ is the black hole mass. We assumed that the metric for smaller radius $r$ preserves its symmetries inherited from the Schwarzschild metric. Namely, it is spherically symmetric and possesses an additional Killing vector, which is spacelike in the black hole interior. Such a metric has 4 independent curvature invariants, which we denoted by $p$, $q$, $u$ and $v$. We restrict our consideration by assuming that the constraint functions are the linear combinations of these invariants.

Certainly, there exists an ambiguity in the  choice  of the coefficients in these linear functions of the invariants and in the number of adopted inequality constraints. In specifying a model we use the following observations.
\begin{itemize}
\item Invariant $v$ coincides with a half of scalar curvature ${}^{(2)}R$ of $2D$ slice spanned by time $\tau$ and Killing vector $\ts{\xi}$ of our 4D metric (see (\ref{gg})). In our previous paper \cite{Frolov:2021kcv} we demonstrated that a restriction imposed on ${}^{(2)}R$ makes the interior of the corresponding 2D black hole free of singularities and the corresponding metric describes an expanding two-dimensional de Sitter universe in the black interior.
    \item Invariants $p$ and $q$ coincide with eigenvalues of the Einstein tensor of 3D slice spanned by time $\tau$ and two-spheres of our 4D metric (see (\ref{GG})). In our previous paper \cite{frolov2021bouncing} we studied in detail linear (as well as more general) constraints, which guarantee the limiting property of corresponding curvature invariants.
\end{itemize}

In the supercritical regime with one inequality constraint the corresponding extended action gives four equations for three functions: $a(\tau)$, $B(\tau)$ and the control function $\chi(\tau)$. Three of them are of the second order in derivatives and one is the first order equation. If the initial conditions satisfy this first order equation, than it is valid for a later time and the system of equations is consistent.
We obtained conditions when  three ``dynamical" equations can be resolved with respect to  the second order derivatives of the field variables. We showed that these conditions are satisfied at the transition point between sub- and supercritical regimes only when the constraints function contains invariant $v$.

Based on these observations we imposed the first constraint of the form $|v|\le \Lambda$. We demonstrated that the invariants $p$ and $q$ are not restricted for such supercritical solution. In order to restrict these invariants, following \cite{frolov2021bouncing} we imposed a secondary constraint in the form $p-\mu q\le \lambda \Lambda$, where $\lambda\ge 1$ and $\mu\in(0,1)$  are dimensionless parameters. After this secondary constraint is saturated, the supercritical solution enters the phase~II, in which both constraints are valid.
Our analysis shows that in this regime all the curvature invariants (including $u$) are restricted and the solution satisfies the limiting curvature condition.

Let us describe main qualitative properties of the solutions for the black hole interior in this model. After the solution enters its phase~II, the metric function $B(\tau)$ continues its growth. The metric function $a(\tau)$, which at beginning of the phase~II is of order of $m\ell$, decreases until it reaches its minimal value $a\approx \ell$. After this it increases until it reaches its maximal value, which is slightly larger than $m\ell$.
Study properties of solutions for control functions associated with constraints shows that, after the solution enters the supercritical regime, it cannot leave it. In other words, the function $a(\tau)$ is periodic. During this periodic motion all the curvature invariants (including $u$) are bounded. Let us note that  similar periodic models for a black hole interior were discussed earlier in \cite{Novikov:1980ni,Muknanov:2017}.

In 1990 Morgan published a paper \cite{Morgan:1990yy}, in which he discussed a model for a black hole interior satisfying a condition of the limiting curvature. He obtained a solution which was quite similar to
the metric proposed earlier in \cite{Frolov:1989pf,Frolov:1988vj}, where a possibility of a new universe formation in the black hole interior was discussed\footnote{ A possibility of several or many new universes creation inside a black hole was discussed in \cite{Barrabes:1995nk}}.
Morgan imposed two constraints, which in our notations are of the form $v=p=\Lambda$, and solved them for the spherically symmetric metric. These equations were chosen ``ad hoc" and he did not derive them from an action. This differs Morgan's approach from the one presented in the present paper. Using the extended action for the inequality constraints allows one not only to obtain the corresponding constraint(s) in the supercritical phase, but also to keep trace of the behavior of the control function(s) which play the role of indicators, informing when transition between different regimes of the solution is possible.
The condition that two phases with $v=\Lambda$ and $p=\Lambda$ starts simultaneously can be achieved in our model as well. However, as we demonstrated, the secondary constraint of the form $p=\Lambda$ is rather singular. For this transition there is a jump of the invariant $q$ at the transition point from  $q=-\frac{1}{2}\Lambda$ to $q=\Lambda$. Such a jump is formally allowed by equations, however it would result in the appearance of a non-integrable singularities  in invariants constructed from covariant derivatives of the curvature, for example
\be
R_{\alpha\beta\gamma\delta;\epsilon}R^{\alpha\beta\gamma\delta;\epsilon}\sim [\delta(\tau)]^2\, .
\ee
This result directly follows from (\ref{DRIN1}). This unpleasant property is absent when $\mu\ne 0$.

When the parameter $\mu$ in the secondary constraint vanishes, the function $a(\tau)$ after bounce still increases, but our results show that, in a general case, it cannot slip to the subcritical solution after this,  since the required conditions for the control functions are not satisfied.
As a result the exponential expansion of both functions $a(\tau)$ and $B(\tau)$ continues forever.
However, as we demonstrated, if one slightly modified the secondary constraint and use the secondary constraint with nonzero $\mu$ instead, the behavior of the function $a(\tau)$ is very different. Namely, it always has a second turning point where it takes maximal value. For small $\mu$ the curvature invariants near this point are large. In this sense the choice of the constraints proposed by Morgan is singular. Periodic property of $a(\tau)$ is valid for any $\mu\in (0,1)$. In order to guarantee that all the invariants are finite and uniformly bounded,
it is sufficient to take $\mu$  to be not too small. We present an explicit solution for the case $\mu=1/2$ which illustrates this property.

As we mentioned, we restricted ourselves by a case of constraints that are linear in curvature invariants. It is interesting to investigate a more general class of constraints.  There are two interesting questions: (i) What happens when one or both constraints are nonlinear functions of basic invariants? (ii) Can one achieve the limiting curvature property by  imposing only one properly chosen inequality constraint?
Partial answer to the first question is the result presented in  \cite{frolov2021bouncing}. Namely, it was shown that one can impose a quite general nonlinear constraint in $(p,q)$ sector of the form $\varphi(p,q,\Lambda)=0$,
which guarantees a similar periodic behavior of $a(\tau)$ as for the linear constraint described in this paper. For this purpose the function $\varphi(p,q,\Lambda)=0$ should satisfy several rather general conditions, discussed in this paper. The second question at the moment is open.

The approach presented in this paper can be used for analysis of the contracting Kasner universe in the limiting gravity model. One can expect that by a proper choice of the inequality curvature constrains one can obtain regular solutions describing bouncing anisotropic universes. This mechanism of suppression of the anisotropy growth in a contracting universe might be of interest in cosmological applications.

\vfill

\appendix

\section{Useful formulas}

One can check that the Ricci tensor $R_{\mu}^{\nu}$ and the Einstein tensor $G_{\mu}^{\nu}$, and the traceless part of the Ricci tensor $S_{\mu}^{\nu}\equiv R_{\mu}^{\nu}-\frac{1}{4}\delta_{\mu}^{\nu}R$ for the metric \eq{metric} can be written as follows
\ba
R_{\mu}^{\nu}&=\mbox{diag}(2q+v, 2u+v, p+q+u, p+q+u) ,\\
-G_{\mu}^{\nu}&=\mbox{diag}(p+2u,p+2q,q+u+v,q+u+v),\\
S_{\mu}^{\nu}&=\frac{1}{2}\mbox{diag}[-p+2q-2u+v, \\
& \hskip1.35cm  -p-2q+2u+v,p-v, p-v].
\ea

We denote
\ba
&{\CAL R}^2=R_{\mu\nu}R^{\mu\nu}, \hskip 1.3cm
{\CAL S}^2=S_{\mu\nu}S^{\mu\nu},\\
&{\CAL C}^2= C_{\alpha\beta\gamma\delta}C^{\alpha\beta\gamma\delta} , \hskip 0.75cm
\cal{K}=R_{\alpha\beta\gamma\delta}R^{\alpha\beta\gamma\delta} .
\ea
Then one has
\bea
&&R=2(p+2q+2u+v), \nonumber \\
&&{\CAL R}^2=(p-v)^2+2(q-u)^2+(p+2q+2u+v)^2 ,\nonumber\\
&&{\CAL S}^2=(p-v)^2+2(q-u)^2,\\
&&{\CAL C}^2=\frac{4}{3}(p-q-u+v)^2 ,\nonumber\\
&&{\cal K}=3(p^2+2q^2+2u^2+v^2) \nonumber  .
\eea

\vfill

\section*{Acknowledgments}

The authors thank the Natural Sciences and Engineering Research Council of Canada and the Killam Trust for their financial support. The authors are grateful to Andrei Frolov for stimulating discussions.
\vfill


\begin{thebibliography}{29}%
\makeatletter
\providecommand \@ifxundefined [1]{%
 \@ifx{#1\undefined}
}%
\providecommand \@ifnum [1]{%
 \ifnum #1\expandafter \@firstoftwo
 \else \expandafter \@secondoftwo
 \fi
}%
\providecommand \@ifx [1]{%
 \ifx #1\expandafter \@firstoftwo
 \else \expandafter \@secondoftwo
 \fi
}%
\providecommand \natexlab [1]{#1}%
\providecommand \enquote  [1]{``#1''}%
\providecommand \bibnamefont  [1]{#1}%
\providecommand \bibfnamefont [1]{#1}%
\providecommand \citenamefont [1]{#1}%
\providecommand \href@noop [0]{\@secondoftwo}%
\providecommand \href [0]{\begingroup \@sanitize@url \@href}%
\providecommand \@href[1]{\@@startlink{#1}\@@href}%
\providecommand \@@href[1]{\endgroup#1\@@endlink}%
\providecommand \@sanitize@url [0]{\catcode `\\12\catcode `\$12\catcode
  `\&12\catcode `\#12\catcode `\^12\catcode `\_12\catcode `\%12\relax}%
\providecommand \@@startlink[1]{}%
\providecommand \@@endlink[0]{}%
\providecommand \url  [0]{\begingroup\@sanitize@url \@url }%
\providecommand \@url [1]{\endgroup\@href {#1}{\urlprefix }}%
\providecommand \urlprefix  [0]{URL }%
\providecommand \Eprint [0]{\href }%
\providecommand \doibase [0]{http://dx.doi.org/}%
\providecommand \selectlanguage [0]{\@gobble}%
\providecommand \bibinfo  [0]{\@secondoftwo}%
\providecommand \bibfield  [0]{\@secondoftwo}%
\providecommand \translation [1]{[#1]}%
\providecommand \BibitemOpen [0]{}%
\providecommand \bibitemStop [0]{}%
\providecommand \bibitemNoStop [0]{.\EOS\space}%
\providecommand \EOS [0]{\spacefactor3000\relax}%
\providecommand \BibitemShut  [1]{\csname bibitem#1\endcsname}%
\let\auto@bib@innerbib\@empty
\bibitem [{\citenamefont {Markov}(1982)}]{Markov:1982}%
  \BibitemOpen
  \bibfield  {author} {\bibinfo {author} {\bibfnamefont {M.A.}\ \bibnamefont
  {Markov}},\ }\bibfield  {title} {\enquote {\bibinfo {title} {{Limiting
  density of matter as a universal law of nature}},}\ }\href@noop {} {\bibfield
   {journal} {\bibinfo  {journal} {JETP Letters}\ }\textbf {\bibinfo {volume}
  {36}},\ \bibinfo {pages} {266} (\bibinfo {year} {1982})}\BibitemShut
  {NoStop}%
\bibitem [{\citenamefont {Markov}(1984)}]{Markov:1984ii}%
  \BibitemOpen
  \bibfield  {author} {\bibinfo {author} {\bibfnamefont {M.A.}\ \bibnamefont
  {Markov}},\ }\bibfield  {title} {\enquote {\bibinfo {title} {{Problems of a
  Perpetually Oscillating Universe}},}\ }\href {\doibase
  10.1016/0003-4916(84)90004-6} {\bibfield  {journal} {\bibinfo  {journal}
  {Annals Phys.}\ }\textbf {\bibinfo {volume} {155}},\ \bibinfo {pages}
  {333--357} (\bibinfo {year} {1984})}\BibitemShut {NoStop}%
\bibitem [{\citenamefont {Gasperini}\ and\ \citenamefont
  {Veneziano}(1993)}]{Gasperini:1992em}%
  \BibitemOpen
  \bibfield  {author} {\bibinfo {author} {\bibfnamefont {M.}~\bibnamefont
  {Gasperini}}\ and\ \bibinfo {author} {\bibfnamefont {G.}~\bibnamefont
  {Veneziano}},\ }\bibfield  {title} {\enquote {\bibinfo {title} {{Pre - big
  bang in string cosmology}},}\ }\href {\doibase 10.1016/0927-6505(93)90017-8}
  {\bibfield  {journal} {\bibinfo  {journal} {Astropart. Phys.}\ }\textbf
  {\bibinfo {volume} {1}},\ \bibinfo {pages} {317--339} (\bibinfo {year}
  {1993})},\ \Eprint {http://arxiv.org/abs/hep-th/9211021}
  {arXiv:hep-th/9211021 [hep-th]} \BibitemShut {NoStop}%
\bibitem [{\citenamefont {Gasperini}\ and\ \citenamefont
  {Veneziano}(2003)}]{Gasperini:2002bn}%
  \BibitemOpen
  \bibfield  {author} {\bibinfo {author} {\bibfnamefont {M.}~\bibnamefont
  {Gasperini}}\ and\ \bibinfo {author} {\bibfnamefont {G.}~\bibnamefont
  {Veneziano}},\ }\bibfield  {title} {\enquote {\bibinfo {title} {{The Pre -
  big bang scenario in string cosmology}},}\ }\href {\doibase
  10.1016/S0370-1573(02)00389-7} {\bibfield  {journal} {\bibinfo  {journal}
  {Phys. Rept.}\ }\textbf {\bibinfo {volume} {373}},\ \bibinfo {pages} {1--212}
  (\bibinfo {year} {2003})},\ \Eprint {http://arxiv.org/abs/hep-th/0207130}
  {arXiv:hep-th/0207130 [hep-th]} \BibitemShut {NoStop}%
\bibitem [{\citenamefont {Mukhanov}\ and\ \citenamefont
  {Brandenberger}(1992)}]{Mukhanov:1991zn}%
  \BibitemOpen
  \bibfield  {author} {\bibinfo {author} {\bibfnamefont {V.~F.}\ \bibnamefont
  {Mukhanov}}\ and\ \bibinfo {author} {\bibfnamefont {R.~H.}\ \bibnamefont
  {Brandenberger}},\ }\bibfield  {title} {\enquote {\bibinfo {title} {{A
  Nonsingular universe}},}\ }\href {\doibase 10.1103/PhysRevLett.68.1969}
  {\bibfield  {journal} {\bibinfo  {journal} {Phys. Rev. Lett.}\ }\textbf
  {\bibinfo {volume} {68}},\ \bibinfo {pages} {1969--1972} (\bibinfo {year}
  {1992})}\BibitemShut {NoStop}%
\bibitem [{\citenamefont {Brandenberger}\ \emph {et~al.}(1993)\citenamefont
  {Brandenberger}, \citenamefont {Mukhanov},\ and\ \citenamefont
  {Sornborger}}]{Brandenberger:1993ef}%
  \BibitemOpen
  \bibfield  {author} {\bibinfo {author} {\bibfnamefont {R.~H.}\ \bibnamefont
  {Brandenberger}}, \bibinfo {author} {\bibfnamefont {V.~F.}\ \bibnamefont
  {Mukhanov}}, \ and\ \bibinfo {author} {\bibfnamefont {A.}~\bibnamefont
  {Sornborger}},\ }\bibfield  {title} {\enquote {\bibinfo {title} {{A
  Cosmological theory without singularities}},}\ }\href {\doibase
  10.1103/PhysRevD.48.1629} {\bibfield  {journal} {\bibinfo  {journal} {Phys.
  Rev. D}\ }\textbf {\bibinfo {volume} {48}},\ \bibinfo {pages} {1629--1642}
  (\bibinfo {year} {1993})},\ \Eprint {http://arxiv.org/abs/gr-qc/9303001}
  {arXiv:gr-qc/9303001} \BibitemShut {NoStop}%
\bibitem [{\citenamefont {Turok}\ and\ \citenamefont
  {Steinhardt}(2005)}]{Turok_2005}%
  \BibitemOpen
  \bibfield  {author} {\bibinfo {author} {\bibfnamefont {N.}~\bibnamefont
  {Turok}}\ and\ \bibinfo {author} {\bibfnamefont {P.J.}\ \bibnamefont
  {Steinhardt}},\ }\bibfield  {title} {\enquote {\bibinfo {title} {Beyond
  inflation a cyclic universe scenario},}\ }\href {\doibase
  10.1238/physica.topical.117a00076} {\bibfield  {journal} {\bibinfo  {journal}
  {Physica Scripta}\ ,\ \bibinfo {pages} {76}} (\bibinfo {year}
  {2005})}\BibitemShut {NoStop}%
\bibitem [{\citenamefont {Biswas}\ \emph {et~al.}(2006)\citenamefont {Biswas},
  \citenamefont {Mazumdar},\ and\ \citenamefont {Siegel}}]{Biswas_2006}%
  \BibitemOpen
  \bibfield  {author} {\bibinfo {author} {\bibfnamefont {T.}~\bibnamefont
  {Biswas}}, \bibinfo {author} {\bibfnamefont {A.}~\bibnamefont {Mazumdar}}, \
  and\ \bibinfo {author} {\bibfnamefont {W.}~\bibnamefont {Siegel}},\
  }\bibfield  {title} {\enquote {\bibinfo {title} {Bouncing universes in
  string-inspired gravity},}\ }\href {\doibase 10.1088/1475-7516/2006/03/009}
  {\bibfield  {journal} {\bibinfo  {journal} {Journal of Cosmology and
  Astroparticle Physics}\ }\textbf {\bibinfo {volume} {2006}},\ \bibinfo
  {pages} {009–009} (\bibinfo {year} {2006})}\BibitemShut {NoStop}%
\bibitem [{\citenamefont {Barvinsky}\ \emph {et~al.}(2008)\citenamefont
  {Barvinsky}, \citenamefont {Kamenshchik},\ and\ \citenamefont
  {Starobinsky}}]{Barvinsky:2008ia}%
  \BibitemOpen
  \bibfield  {author} {\bibinfo {author} {\bibfnamefont {A.~O.}\ \bibnamefont
  {Barvinsky}}, \bibinfo {author} {\bibfnamefont {A.~Yu.}\ \bibnamefont
  {Kamenshchik}}, \ and\ \bibinfo {author} {\bibfnamefont {A.~A.}\ \bibnamefont
  {Starobinsky}},\ }\bibfield  {title} {\enquote {\bibinfo {title} {{Inflation
  scenario via the Standard Model Higgs boson and LHC}},}\ }\href {\doibase
  10.1088/1475-7516/2008/11/021} {\bibfield  {journal} {\bibinfo  {journal}
  {JCAP}\ }\textbf {\bibinfo {volume} {11}},\ \bibinfo {pages} {021} (\bibinfo
  {year} {2008})},\ \Eprint {http://arxiv.org/abs/0809.2104} {arXiv:0809.2104
  [hep-ph]} \BibitemShut {NoStop}%
\bibitem [{\citenamefont {Novello}\ and\ \citenamefont
  {Bergliaffa}(2008)}]{Novello:2008ra}%
  \BibitemOpen
  \bibfield  {author} {\bibinfo {author} {\bibfnamefont {M.}~\bibnamefont
  {Novello}}\ and\ \bibinfo {author} {\bibfnamefont {S.~E.~Perez}\ \bibnamefont
  {Bergliaffa}},\ }\bibfield  {title} {\enquote {\bibinfo {title} {{Bouncing
  Cosmologies}},}\ }\href {\doibase 10.1016/j.physrep.2008.04.006} {\bibfield
  {journal} {\bibinfo  {journal} {Phys. Rept.}\ }\textbf {\bibinfo {volume}
  {463}},\ \bibinfo {pages} {127--213} (\bibinfo {year} {2008})},\ \Eprint
  {http://arxiv.org/abs/0802.1634} {arXiv:0802.1634 [astro-ph]} \BibitemShut
  {NoStop}%
\bibitem [{\citenamefont {Lehners}(2008)}]{Lehners:2008vx}%
  \BibitemOpen
  \bibfield  {author} {\bibinfo {author} {\bibfnamefont {J.-L.}\ \bibnamefont
  {Lehners}},\ }\bibfield  {title} {\enquote {\bibinfo {title} {{Ekpyrotic and
  Cyclic Cosmology}},}\ }\href {\doibase 10.1016/j.physrep.2008.06.001}
  {\bibfield  {journal} {\bibinfo  {journal} {Phys. Rept.}\ }\textbf {\bibinfo
  {volume} {465}},\ \bibinfo {pages} {223--263} (\bibinfo {year} {2008})},\
  \Eprint {http://arxiv.org/abs/0806.1245} {arXiv:0806.1245 [astro-ph]}
  \BibitemShut {NoStop}%
\bibitem [{\citenamefont {Ashtekar}(2009)}]{Ashtekar_2009}%
  \BibitemOpen
  \bibfield  {author} {\bibinfo {author} {\bibfnamefont {A.}~\bibnamefont
  {Ashtekar}},\ }\bibfield  {title} {\enquote {\bibinfo {title} {Singularity
  resolution in loop quantum cosmology: A brief overview},}\ }\href {\doibase
  10.1088/1742-6596/189/1/012003} {\bibfield  {journal} {\bibinfo  {journal}
  {Journal of Physics: Conference Series}\ }\textbf {\bibinfo {volume} {189}},\
  \bibinfo {pages} {012003} (\bibinfo {year} {2009})}\BibitemShut {NoStop}%
\bibitem [{\citenamefont {Cesar~e Silva}\ and\ \citenamefont
  {Shapiro}(2020)}]{CesareSilva:2020ihf}%
  \BibitemOpen
  \bibfield  {author} {\bibinfo {author} {\bibfnamefont {W.}~\bibnamefont
  {Cesar~e Silva}}\ and\ \bibinfo {author} {\bibfnamefont {I.~L.}\ \bibnamefont
  {Shapiro}},\ }\bibfield  {title} {\enquote {\bibinfo {title} {{Bounce and
  Stability in the Early Cosmology with Anomaly-Induced Corrections}},}\ }\href
  {\doibase 10.3390/sym13010050} {\bibfield  {journal} {\bibinfo  {journal}
  {Symmetry}\ }\textbf {\bibinfo {volume} {13}},\ \bibinfo {pages} {50}
  (\bibinfo {year} {2020})},\ \Eprint {http://arxiv.org/abs/2012.10554}
  {arXiv:2012.10554 [hep-th]} \BibitemShut {NoStop}%
\bibitem [{\citenamefont {Biswas}\ \emph {et~al.}(2012)\citenamefont {Biswas},
  \citenamefont {Koshelev}, \citenamefont {Mazumdar},\ and\ \citenamefont
  {Vernov}}]{Biswas:2012bp}%
  \BibitemOpen
  \bibfield  {author} {\bibinfo {author} {\bibfnamefont {T.}~\bibnamefont
  {Biswas}}, \bibinfo {author} {\bibfnamefont {A.~S.}\ \bibnamefont
  {Koshelev}}, \bibinfo {author} {\bibfnamefont {A.}~\bibnamefont {Mazumdar}},
  \ and\ \bibinfo {author} {\bibfnamefont {S.~Yu.}\ \bibnamefont {Vernov}},\
  }\bibfield  {title} {\enquote {\bibinfo {title} {{Stable bounce and inflation
  in non-local higher derivative cosmology}},}\ }\href {\doibase
  10.1088/1475-7516/2012/08/024} {\bibfield  {journal} {\bibinfo  {journal}
  {JCAP}\ }\textbf {\bibinfo {volume} {08}},\ \bibinfo {pages} {024} (\bibinfo
  {year} {2012})},\ \Eprint {http://arxiv.org/abs/1206.6374} {arXiv:1206.6374
  [astro-ph.CO]} \BibitemShut {NoStop}%
\bibitem [{\citenamefont {Battefeld}\ and\ \citenamefont
  {Peter}(2015)}]{Battefeld:2014uga}%
  \BibitemOpen
  \bibfield  {author} {\bibinfo {author} {\bibfnamefont {D.}~\bibnamefont
  {Battefeld}}\ and\ \bibinfo {author} {\bibfnamefont {P.}~\bibnamefont
  {Peter}},\ }\bibfield  {title} {\enquote {\bibinfo {title} {{A Critical
  Review of Classical Bouncing Cosmologies}},}\ }\href {\doibase
  10.1016/j.physrep.2014.12.004} {\bibfield  {journal} {\bibinfo  {journal}
  {Phys. Rept.}\ }\textbf {\bibinfo {volume} {571}},\ \bibinfo {pages} {1--66}
  (\bibinfo {year} {2015})},\ \Eprint {http://arxiv.org/abs/1406.2790}
  {arXiv:1406.2790 [astro-ph.CO]} \BibitemShut {NoStop}%
\bibitem [{\citenamefont {Brandenberger}\ and\ \citenamefont
  {Peter}(2017)}]{Brandenberger_2017}%
  \BibitemOpen
  \bibfield  {author} {\bibinfo {author} {\bibfnamefont {R.}~\bibnamefont
  {Brandenberger}}\ and\ \bibinfo {author} {\bibfnamefont {P.}~\bibnamefont
  {Peter}},\ }\bibfield  {title} {\enquote {\bibinfo {title} {Bouncing
  cosmologies: Progress and problems},}\ }\href {\doibase
  10.1007/s10701-016-0057-0} {\bibfield  {journal} {\bibinfo  {journal}
  {Foundations of Physics}\ }\textbf {\bibinfo {volume} {47}},\ \bibinfo
  {pages} {797–850} (\bibinfo {year} {2017})}\BibitemShut {NoStop}%
\bibitem [{\citenamefont {Yoshida}\ \emph {et~al.}(2017)\citenamefont
  {Yoshida}, \citenamefont {Quintin}, \citenamefont {Yamaguchi},\ and\
  \citenamefont {Brandenberger}}]{Yoshida:2017swb}%
  \BibitemOpen
  \bibfield  {author} {\bibinfo {author} {\bibfnamefont {D.}~\bibnamefont
  {Yoshida}}, \bibinfo {author} {\bibfnamefont {J.}~\bibnamefont {Quintin}},
  \bibinfo {author} {\bibfnamefont {M.}~\bibnamefont {Yamaguchi}}, \ and\
  \bibinfo {author} {\bibfnamefont {R.~H.}\ \bibnamefont {Brandenberger}},\
  }\bibfield  {title} {\enquote {\bibinfo {title} {{Cosmological perturbations
  and stability of nonsingular cosmologies with limiting curvature}},}\ }\href
  {\doibase 10.1103/PhysRevD.96.043502} {\bibfield  {journal} {\bibinfo
  {journal} {Phys. Rev. D}\ }\textbf {\bibinfo {volume} {96}},\ \bibinfo
  {pages} {043502} (\bibinfo {year} {2017})},\ \Eprint
  {http://arxiv.org/abs/1704.04184} {arXiv:1704.04184 [hep-th]} \BibitemShut
  {NoStop}%
\bibitem [{\citenamefont {Ijjas}\ and\ \citenamefont
  {Steinhardt}(2018)}]{Ijjas_2018}%
  \BibitemOpen
  \bibfield  {author} {\bibinfo {author} {\bibfnamefont {A.}~\bibnamefont
  {Ijjas}}\ and\ \bibinfo {author} {\bibfnamefont {P.J.}\ \bibnamefont
  {Steinhardt}},\ }\bibfield  {title} {\enquote {\bibinfo {title} {Bouncing
  cosmology made simple},}\ }\href {\doibase 10.1088/1361-6382/aac482}
  {\bibfield  {journal} {\bibinfo  {journal} {Classical and Quantum Gravity}\
  }\textbf {\bibinfo {volume} {35}},\ \bibinfo {pages} {135004} (\bibinfo
  {year} {2018})}\BibitemShut {NoStop}%
\bibitem [{\citenamefont {Frolov}\ \emph {et~al.}(1990)\citenamefont {Frolov},
  \citenamefont {Markov},\ and\ \citenamefont {Mukhanov}}]{Frolov:1988vj}%
  \BibitemOpen
  \bibfield  {author} {\bibinfo {author} {\bibfnamefont {V.P.}\ \bibnamefont
  {Frolov}}, \bibinfo {author} {\bibfnamefont {M.A.}\ \bibnamefont {Markov}}, \
  and\ \bibinfo {author} {\bibfnamefont {V.F.}\ \bibnamefont {Mukhanov}},\
  }\bibfield  {title} {\enquote {\bibinfo {title} {{Black Holes as Possible
  Sources of Closed and Semiclosed Worlds}},}\ }\href {\doibase
  10.1103/PhysRevD.41.383} {\bibfield  {journal} {\bibinfo  {journal} {Phys.
  Rev.}\ }\textbf {\bibinfo {volume} {D41}},\ \bibinfo {pages} {383} (\bibinfo
  {year} {1990})}\BibitemShut {NoStop}%
\bibitem [{\citenamefont {Frolov}\ \emph {et~al.}(1989)\citenamefont {Frolov},
  \citenamefont {Markov},\ and\ \citenamefont {Mukhanov}}]{Frolov:1989pf}%
  \BibitemOpen
  \bibfield  {author} {\bibinfo {author} {\bibfnamefont {V.P.}\ \bibnamefont
  {Frolov}}, \bibinfo {author} {\bibfnamefont {M.A.}\ \bibnamefont {Markov}}, \
  and\ \bibinfo {author} {\bibfnamefont {V.F.}\ \bibnamefont {Mukhanov}},\
  }\bibfield  {title} {\enquote {\bibinfo {title} {{Through a black hole into a
  new universe?}}}\ }\href {\doibase 10.1016/0370-2693(89)91114-3} {\bibfield
  {journal} {\bibinfo  {journal} {Phys. Lett.}\ }\textbf {\bibinfo {volume}
  {B216}},\ \bibinfo {pages} {272--276} (\bibinfo {year} {1989})}\BibitemShut
  {NoStop}%
\bibitem [{\citenamefont {Morgan}(1991)}]{Morgan:1990yy}%
  \BibitemOpen
  \bibfield  {author} {\bibinfo {author} {\bibfnamefont {D.}~\bibnamefont
  {Morgan}},\ }\bibfield  {title} {\enquote {\bibinfo {title} {{Black holes in
  cutoff gravity}},}\ }\href {\doibase 10.1103/PhysRevD.43.3144} {\bibfield
  {journal} {\bibinfo  {journal} {Phys. Rev. D}\ }\textbf {\bibinfo {volume}
  {43}},\ \bibinfo {pages} {3144--3146} (\bibinfo {year} {1991})}\BibitemShut
  {NoStop}%
\bibitem [{\citenamefont {Barrabes}\ and\ \citenamefont
  {Frolov}(1996)}]{Barrabes:1995nk}%
  \BibitemOpen
  \bibfield  {author} {\bibinfo {author} {\bibfnamefont {C.}~\bibnamefont
  {Barrabes}}\ and\ \bibinfo {author} {\bibfnamefont {V.~P.}\ \bibnamefont
  {Frolov}},\ }\bibfield  {title} {\enquote {\bibinfo {title} {{How many new
  worlds are inside a black hole?}}}\ }\href {\doibase
  10.1103/PhysRevD.53.3215} {\bibfield  {journal} {\bibinfo  {journal} {Phys.
  Rev. D}\ }\textbf {\bibinfo {volume} {53}},\ \bibinfo {pages} {3215--3223}
  (\bibinfo {year} {1996})},\ \Eprint {http://arxiv.org/abs/hep-th/9511136}
  {arXiv:hep-th/9511136} \BibitemShut {NoStop}%
\bibitem [{\citenamefont {Chamseddine}\ and\ \citenamefont
  {Mukhanov}(2017)}]{Muknanov:2017}%
  \BibitemOpen
  \bibfield  {author} {\bibinfo {author} {\bibfnamefont {A.~H.}\ \bibnamefont
  {Chamseddine}}\ and\ \bibinfo {author} {\bibfnamefont {V.}~\bibnamefont
  {Mukhanov}},\ }\bibfield  {title} {\enquote {\bibinfo {title} {Nonsingular
  black hole},}\ }\href {\doibase 10.1140/epjc/s10052-017-4759-z} {\bibfield
  {journal} {\bibinfo  {journal} {The European Physical Journal C}\ }\textbf
  {\bibinfo {volume} {77}} (\bibinfo {year} {2017}),\
  10.1140/epjc/s10052-017-4759-z}\BibitemShut {NoStop}%
\bibitem [{\citenamefont {Frolov}\ and\ \citenamefont {Frolov}()}]{AVF}%
  \BibitemOpen
  \bibfield  {author} {\bibinfo {author} {\bibfnamefont {Andrei~V.}\
  \bibnamefont {Frolov}}\ and\ \bibinfo {author} {\bibfnamefont {Valeri~P.}\
  \bibnamefont {Frolov}},\ }\bibfield  {title} {\enquote {\bibinfo {title}
  {{Classical mechanics with inequality constraints}},}\ }\href@noop {} {\
  }\BibitemShut {NoStop}%
\bibitem [{\citenamefont {Frolov}\ and\ \citenamefont
  {Zelnikov}(2021{\natexlab{a}})}]{Frolov:2021kcv}%
  \BibitemOpen
  \bibfield  {author} {\bibinfo {author} {\bibfnamefont {V.~P.}\ \bibnamefont
  {Frolov}}\ and\ \bibinfo {author} {\bibfnamefont {A.}~\bibnamefont
  {Zelnikov}},\ }\bibfield  {title} {\enquote {\bibinfo {title}
  {{Two-dimensional black holes in the limiting curvature theory of
  gravity}},}\ }\href@noop {} {\  (\bibinfo {year} {2021}{\natexlab{a}})},\
  \Eprint {http://arxiv.org/abs/2105.12808} {arXiv:2105.12808 [hep-th]}
  \BibitemShut {NoStop}%
\bibitem [{\citenamefont {Frolov}\ and\ \citenamefont
  {Zelnikov}(2021{\natexlab{b}})}]{frolov2021bouncing}%
  \BibitemOpen
  \bibfield  {author} {\bibinfo {author} {\bibfnamefont {V.~P.}\ \bibnamefont
  {Frolov}}\ and\ \bibinfo {author} {\bibfnamefont {A.}~\bibnamefont
  {Zelnikov}},\ }\href@noop {} {\enquote {\bibinfo {title} {Bouncing cosmology
  in the limiting curvature theory of gravity},}\ } (\bibinfo {year}
  {2021}{\natexlab{b}}),\ \Eprint {http://arxiv.org/abs/2108.09927}
  {arXiv:2108.09927 [hep-th]} \BibitemShut {NoStop}%
\bibitem [{\citenamefont {Narlikar}\ and\ \citenamefont
  {Karmarkar}(1949)}]{NARLIKAR}%
  \BibitemOpen
  \bibfield  {author} {\bibinfo {author} {\bibfnamefont {V.~V.}\ \bibnamefont
  {Narlikar}}\ and\ \bibinfo {author} {\bibfnamefont {K.~R.}\ \bibnamefont
  {Karmarkar}},\ }\bibfield  {title} {\enquote {\bibinfo {title} {{The scalar
  invariants of a general gravitational metric}},}\ }\href@noop {} {\bibfield
  {journal} {\bibinfo  {journal} {Proceedings of the Indian Academy of
  Sciences, Section A}\ }\textbf {\bibinfo {volume} {29}},\ \bibinfo {pages}
  {91--97} (\bibinfo {year} {1949})}\BibitemShut {NoStop}%
\bibitem [{\citenamefont {Santosuosso}\ \emph {et~al.}(1998)\citenamefont
  {Santosuosso}, \citenamefont {Pollney}, \citenamefont {Pelavas},
  \citenamefont {Musgrave},\ and\ \citenamefont {Lake}}]{CURV_INV_1998}%
  \BibitemOpen
  \bibfield  {author} {\bibinfo {author} {\bibfnamefont {K.}~\bibnamefont
  {Santosuosso}}, \bibinfo {author} {\bibfnamefont {D.}~\bibnamefont
  {Pollney}}, \bibinfo {author} {\bibfnamefont {N.}~\bibnamefont {Pelavas}},
  \bibinfo {author} {\bibfnamefont {P.}~\bibnamefont {Musgrave}}, \ and\
  \bibinfo {author} {\bibfnamefont {K.}~\bibnamefont {Lake}},\ }\bibfield
  {title} {\enquote {\bibinfo {title} {Invariants of the riemann tensor for
  class b warped product space-times},}\ }\href@noop {} {\bibfield  {journal}
  {\bibinfo  {journal} {Computer Physics Communications}\ }\textbf {\bibinfo
  {volume} {115}},\ \bibinfo {pages} {381–394} (\bibinfo {year}
  {1998})}\BibitemShut {NoStop}%
\bibitem [{\citenamefont {Novikov}\ and\ \citenamefont
  {Starobinsky}(1980)}]{Novikov:1980ni}%
  \BibitemOpen
  \bibfield  {author} {\bibinfo {author} {\bibfnamefont {I.~D.}\ \bibnamefont
  {Novikov}}\ and\ \bibinfo {author} {\bibfnamefont {A.~A.}\ \bibnamefont
  {Starobinsky}},\ }\bibfield  {title} {\enquote {\bibinfo {title} {{Quantum
  electrodynamic effects inside a charged black hole and the problem of Cauchy
  horizons}},}\ }\href@noop {} {\bibfield  {journal} {\bibinfo  {journal} {Sov.
  Phys. JETP}\ }\textbf {\bibinfo {volume} {51}},\ \bibinfo {pages} {1--9}
  (\bibinfo {year} {1980})}\BibitemShut {NoStop}%
\end{thebibliography}

%

\end{document}